\documentclass[aps,prc,amsmath,amssymb,twocolumn,superscriptaddress,showpacs,floatfix]{revtex4}

\usepackage[dvips]{graphicx}
\usepackage[usenames]{color}
\usepackage{placeins}

\begin{document}
\newcommand{\gevc}{\rm(GeV/c)$^2$~}
\newcommand{\gevcp}{\rm(GeV/c)$^2$}
\definecolor{MyHQM}{rgb}{0.57,0,1}
\definecolor{MyCapstick}{rgb}{0,0.7,0.7}


\title{Measurements of the
  $\gamma^*p\rightarrow\Delta$ Reaction At Low $Q^2$: Probing the
  Mesonic Contribution}


\newcommand{\mitlns}{Department of Physics, Laboratory
for Nuclear Science and Bates Linear Accelerator Center, Massachusetts Institute of Technology, Cambridge,
Massachusetts 02139, USA}

\newcommand{\mitlnscurr}{Current address: Department of Physics, Laboratory
for Nuclear Science and Bates Linear Accelerator Center, Massachusetts Institute of Technology, Cambridge,
Massachusetts 02139, USA}

\newcommand{\mainz}{Institute f\"ur Kernphysik, Johannes Gutenberg-Universit\"at Mainz, D-55099  Mainz, Germany}

\newcommand{\duke}{Current address: Triangle Universities Nuclear Laboratory, Duke University, Durham, North Carolina 27708, USA}

\newcommand{\riken}{Current address: Radiation Laboratory, RIKEN, 2-1 Hirosawa, Wako, Saitama 351-0198, Japan}

\newcommand{\bonn} {Current address: Univ. Bonn, Physikalisches Inst.,
Nussallee 12, D-53115 Bonn, Germany}

\newcommand{\athens}{Institute of Accelerating Systems and Applications and
  Department of Physics, University of Athens, Athens, Greece}

\newcommand{\zagreb}{Department of Physics, University of Zagreb, Croatia}

\newcommand{\uk}{Department of Physics and Astronomy, University of
  Kentucky, Lexington, Kentucky 40206 USA}

\newcommand{\ulj}{Institute Jo\v zef Stefan, University of Ljubljana,
  Ljubljana, Slovenia}

\author{S. Stave}\altaffiliation{\duke}\affiliation{\mitlns}

\author{N. Sparveris}\altaffiliation{\mitlnscurr}\affiliation{\athens}
\author{M.~O.~Distler}\affiliation{\mainz}
\author{I.~Nakagawa}\altaffiliation{\riken}\affiliation{\mitlns}\affiliation{\uk}
\author{P.~Achenbach}\affiliation{\mainz}
\author{C.~Ayerbe Gayoso}\affiliation{\mainz}
\author{D.~Baumann}\affiliation{\mainz}
\author{J.~Bernauer}\affiliation{\mainz}
\author{A. M. Bernstein}\altaffiliation{Corresponding author}\affiliation{\mitlns}
\author{R.~B\"ohm}\affiliation{\mainz}
\author{D.~Bosnar}\affiliation{\zagreb}
\author{T. Botto}\affiliation{\mitlns}
\author{A.~Christopoulou}\affiliation{\athens}
\author{D.~Dale}\affiliation{\uk}
\author{M.~Ding}\affiliation{\mainz}
\author{L.~Doria}\affiliation{\mainz}
\author{J.~Friedrich}\affiliation{\mainz}
\author{A.~Karabarbounis}\affiliation{\athens}
\author{M.~Makek}\affiliation{\zagreb}
\author{H.~Merkel}\affiliation{\mainz}
\author{U.~M\"uller}\affiliation{\mainz}
\author{R.~Neuhausen}\affiliation{\mainz}
\author{L.~Nungesser}\affiliation{\mainz}
\author{C.N.~Papanicolas}\affiliation{\athens}
\author{A.~Piegsa}\affiliation{\mainz}
\author{J.~Pochodzalla}\affiliation{\mainz}
\author{M.~Potokar}\affiliation{\ulj}
\author{M.~Seimetz}\altaffiliation{\bonn}\affiliation{\mainz}
\author{S.~\v Sirca}\affiliation{\ulj}
\author{S.~Stiliaris}\affiliation{\athens}
\author{Th.~Walcher}\affiliation{\mainz}
\author{M.~Weis}\affiliation{\mainz}
\collaboration{A1 Collaboration}\noaffiliation


\date{\today}

\begin{abstract}
The determination of non-spherical angular momentum amplitudes in
nucleons at long ranges (low $Q^{2}$), was accomplished through
the $p(\vec{e},e'p)\pi^0$ reaction in the $\Delta$ region at
$Q^2=0.060$, 0.127, and 0.200 \gevc at the Mainz Microtron (MAMI) with an
accuracy of 3\%.
The results for the dominant
transition magnetic dipole amplitude  and the quadrupole to dipole
ratios have been obtained with an estimated model uncertainty  which is
approximately the same as the experimental uncertainty.
Lattice and  effective field theory predictions agree
with our  data within the relatively large estimated
theoretical uncertainties.
Phenomenological models
are in good agreement with experiment when the
resonant amplitudes are adjusted to the data. To check
reaction model calculations additional
data were taken for center of mass energies below resonance and for the
$\sigma_{TL'}$ structure function.  These results confirm the dominance, and
general $Q^{2}$
variation, of the  pionic contribution at large distances.
\end{abstract}


\maketitle


\section{Introduction}

Experimental confirmation of the presence of non-spherical hadron amplitudes (i.e. d states in quark models or p wave $\pi$-N states) is fundamental and has been the subject of intense experimental and theoretical interest (for reviews see~\cite{athens2006,mit2004,nstar2001,cnp,amb}). This effort has focused on the measurement of the electric and Coulomb quadrupole amplitudes (E2, C2) in the predominantly M1 (magnetic dipole-quark spin flip) $\gamma^* N\rightarrow \Delta$ transition.

The present low $Q^2$ experiments add important data to determine the
physical basis of long range nucleon and $\Delta$ non-spherical
amplitudes. This is the region where pionic effects are predicted to
be dominant and appreciably changing.  The experiment was carried out at
the Mainz Microtron to measure cross sections and extract the resonant
multipoles at $Q^2=0.060$, 0.127, and 0.200 \gevcp.  The $Q^2 = 0.060$ \gevc point is the lowest $Q^{2}$ value probed to date in
modern electroproduction experiments.  The
$Q^2=0.200$ \gevc point tests the $Q^2$ variation and provides a
valuable overlap with newly obtained Jefferson Lab data~\cite{CLAS2007}.  The $Q^2=0.127$ \gevc point
tested the background amplitudes and acted as a comparison of results
from Mainz and Bates.  Aspects of this work are given in
\cite{stave,sparveris_Q20}.  This work includes more of the details
and is a complete account of those data and includes previously
unpublished data as well.

The present measurements fill in an important gap in the coverage of
the $Q^2$ evolution between the photon
point($Q^2=0$)~\cite{beck,blanpied} and previously published
electroproduction experiments at JLab~\cite{joo,frolov,ungaro} for $Q^2$ from
0.4 to 6.0 (GeV/c)$^2$, with the exception of good coverage at $Q^2$ =
0.127 \gevc at Bates~\cite{warren,mertz,kunz,sparveris}  and
Mainz~\cite{pospischil,bartsch, elsner} at $Q^2$ = 0.127, 0.200
(GeV/c)$^2$ that have been published. 
 
Since the proton has spin 1/2, no quadrupole moment can be
measured.  However, the $\Delta$ has spin 3/2 so the 
$\gamma^*N\rightarrow \Delta$ reaction can be studied for quadrupole amplitudes in the nucleon and $\Delta$.  
Due to spin and parity conservation in the $\gamma^*N(J^\pi=1/2^+)
\rightarrow \Delta(J^\pi=3/2^+)$ reaction, only three multipoles can
contribute to the transition: the magnetic dipole ($M1$), the electric
quadrupole ($E2$), and the Coulomb quadrupole ($C2$) photon absorption
multipoles.  The corresponding resonant pion production multipoles are 
$M_{1+}^{3/2}$, $E_{1+}^{3/2}$, and $S_{1+}^{3/2}$.  The relative quadrupole to dipole ratios are
EMR=Re($E_{1+}^{3/2}/M_{1+}^{3/2}$) and 
CMR=Re($S_{1+}^{3/2}/M_{1+}^{3/2}$).
In the quark model, the non-spherical amplitudes in the 
nucleon and $\Delta$ are caused by the non-central, tensor interaction
between quarks
~\cite{glashow,isgur_karl}.
However, the magnitudes of this effect for the predicted E2 and C2  amplitudes~\cite{capstick_karl} are at least an
order of magnitude too small to explain the experimental results (see
Fig.~\ref{fig:mpole_vs_Q2} below) and
even the dominant M1 matrix element is $\simeq$ 30\% low~\cite{amb,capstick_karl}. A likely cause of these dynamical shortcomings is that the quark model
does not respect chiral symmetry, whose spontaneous breaking leads to strong emission of virtual pions (Nambu-Goldstone Bosons)~\cite{amb}. These couple
to  nucleons as $\vec{\sigma}\cdot \vec{p}$ where  $\vec{\sigma}$ is the nucleon spin, and $\vec{p}$ is the pion momentum. The coupling is strong in the p wave and mixes in non-zero
angular momentum components. Based on this,
it is physically reasonable to expect that the pionic contributions increase the M1  and 
dominate the E2 and C2 transition matrix elements in the low $Q^2$
(large distance) domain. This was first indicated by adding pionic effects to quark models~\cite{quark_pion_1,quark_pion_2,quark_pion_3}, subsequently in pion cloud model calculations~\cite{sato_lee,dmt}, and 
recently demonstrated in  effective field theory (chiral) calculations~\cite{gail_hemmert,pasc}.

\section{Equipment}

The $p(\vec{e},e'p)\pi^{0}$ measurements were performed using the A1
spectrometers at the Mainz Microtron~\cite{blom}.  
 Electrons were
detected in Spectrometer A which used two pairs of vertical drift
chambers for track reconstruction and two layers of scintillator
detectors for timing information and particle identification.  
 The protons were
detected in Spectrometer B which has a detector package similar to
Spectrometer A.  Spectrometer B also has the ability to measure at up
to 10$^\circ$ out-of-plane in the lab.  Due to the Lorentz boost, 
this corresponds to a significantly
larger value in the center of mass frame.  The momentum resolution of the spectrometers is 0.01\% and the angular
resolution at the target is 3 mrad~\cite{blom}.  Details about the
spectrometers are available in~\cite{blom}.
The MAMI B accelerator delivered a longitudinally polarized,
continuous, electron beam up to 855 MeV. Beam polarization
was measured periodically with a M\o{}ller polarimeter~\cite{bartsch_thesis} to be $\approx
75\%$. The beam with average current of up to 25 $\mu$A was
scattered from a liquid hydrogen cryogenic target. 
The beam energy has an absolute uncertainty of $\pm 160$ keV and a spread of 30 keV (FWHM)~\cite{blom}.  The effects of these uncertainties and the various kinematic cuts were studied  to estimate an overall
systematic uncertainty (see Table \ref{table:sys_errs}) 
for the cross sections  of 3 to 4\%. 
This was tested with elastic electron-proton scattering and the data agree with a fit to the world data~\cite{mergell} at the 3\% level.  
In addition, a third spectrometer (Spectrometer C) 
was used throughout the experiment as
a luminosity monitor.


\section{Experimental methodology}


The five-fold differential cross section for the
$p(\vec{e},e'p)\pi^{0}$ reaction is written as five two-fold
differential cross sections with an explicit $\phi^*$
dependence as~\cite{drechsel_tiator}

\begin{eqnarray}
\displaystyle \frac{d^5\sigma}{d\Omega_f dE_f d\Omega}  = & \Gamma (\sigma_T + \epsilon
\sigma_L + v_{LT}\sigma_{LT}\cos\phi_{\pi q}^* \nonumber \\ 
 + & \displaystyle \epsilon \sigma_{TT} \cos 2\phi_{\pi q}^* 
 +  h p_e v_{LT'}\sigma_{LT'} \sin \phi_{\pi q}^*)
\label{eq:XS}
\end{eqnarray}

where $\phi_{\pi q}^*$ is the pion center of mass azimuthal angle with
respect to the electron scattering plane, $h$ is the 
helicity of the electron beam, $p_e$ is the polarization of the
electron beam, $v_{LT}=\sqrt{2\epsilon(1+\epsilon)}$, $v_{LT'}=\sqrt{2\epsilon(1-\epsilon)}$, $\epsilon$ is the transverse polarization of the virtual photon,
and $\Gamma$ is the virtual photon flux.
The virtual photon differential cross sections
($\sigma_{T},\sigma_{L},\sigma_{LT},\sigma_{TT},\sigma_{LT'}$) 
are all functions of the
center of mass energy $W$, the four momentum transfer squared $Q^2$,
and the pion center of mass polar angle $\theta_{\pi q}^{*}$ (measured
from the momentum transfer direction). They are bilinear combinations
of the multipoles~\cite{drechsel_tiator}.

The extraction of the cross sections was performed using three
sequential measurements. For the helicity independent cross sections there are
three cross sections to extract: $\sigma_0=\sigma_T+\epsilon\sigma_L$
($\epsilon$ was not varied so the two cross sections cannot be separated),
$\sigma_{TT}$ and $\sigma_{LT}$.  The three two-fold differential 
cross sections can be
extracted algebraically 
by measuring the five-fold differential cross section at the same
center-of-mass energy $W$, four-momentum transfer squared $Q^2$, and
proton center-of-mass polar angle $\theta_{pq}^*$ but different values
of the proton azimuthal angle $\phi_{pq}^*$.
(The proton and pion are back-to-back in the center of mass frame
leading to the following relations between the angles:
$\theta_{pq}^*=180^\circ-\theta_{\pi q}^*$ and $\phi_{pq}^*= 180^\circ
+ \phi_{\pi q}^*$.)
The sequential kinematic settings then keep $W$ and $Q^2$ constant by
keeping the electron arm (Spectrometer A) unchanged and the proton arm
was moved so that $\theta_{pq}^*$ remained the same and $\phi_{pq}^*$
was changed.  Using Eq.~\ref{eq:XS}, 
the three two-fold differential cross sections ($\sigma_0,
\sigma_{TT}, \sigma_{LT}$) can then be found algebraically from the
three measured five-fold cross sections and the $\phi_{pq}^{*}$
angles at which they were measured.
The fifth two-fold differential cross section 
$\sigma_{LT'}$ was measured by 
reversing the helicity of the 
longitudinally  polarized electron beam
at non-zero (out-of-plane) $\phi_{pq}^*$ angles.  $\sigma_{LT'}$ is
sensitive to the background terms 
and provides another test of the reaction calculations.

\begin{figure*}
\includegraphics[angle=0,width=1.6\columnwidth]{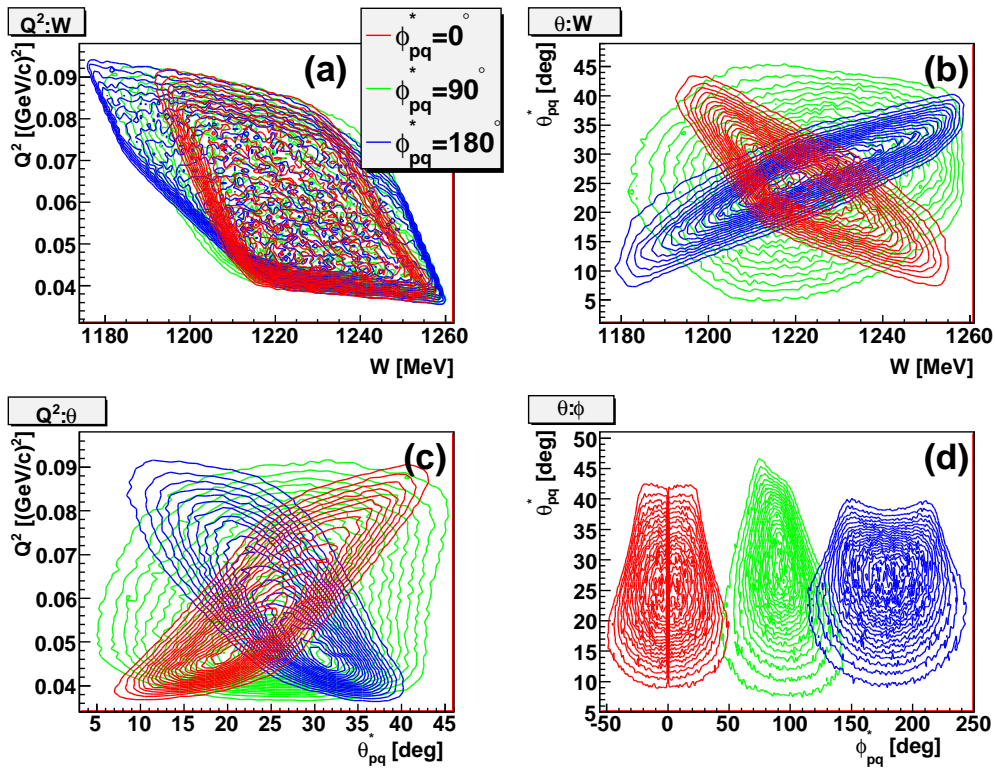}
\caption{\label{fig:acc_overlap}(Color online) Plot of the overlap of the sequential settings
  for $Q^2=0.060$ \gevcp,$W=1221$ MeV, $\theta_{pq}^*=24^\circ$.
  Medium gray corresponds to $\phi_{pq}^{*}=0^\circ$, light gray to
  $\phi_{pq}^{*}=90^\circ$ and dark gray to
  $\phi_{pq}^{*}=180^\circ$.  The amount of overlap for $Q^2=0.20$
  \gevc is qualitatively similar.}
\end{figure*}

Figure \ref{fig:acc_overlap} shows the
kinematic overlap for the sequential $\phi_{pq}^*$ settings at
$Q^2=0.060$ \gevcp.
The $W$ overlap is approximately 40 MeV, the
$\Delta Q^2\approx 0.04$ GeV$^2$/c$^2$, $\Delta \theta_{pq}^{*}\approx
10^\circ$, and $\Delta\phi_{pq}^{*}\approx 40^\circ$.  For $Q^2=0.200$
\gevcp, the overlap region is slightly larger due to the larger
Lorentz boost but the shapes are qualitatively similar.

Studies of the extraction process showed that the smallest uncertainties and
most sensitivity were achieved when the three $\phi_{pq}^*$
measurements were as far apart as possible.  However, at the larger
$\theta_{pq}^*$ angles, not all $\phi_{pq}^*$ values can be reached
because of the $10^\circ$ out-of-plane angle constraint of Spectrometer
B.  Therefore, for each $\theta_{pq}^*$ setting, the maximally
out-of-plane settings were used.  However, each kinematic setting was
carefully chosen in order to minimize the uncertainties in the
algebraic cross section extraction process.

The kinematics for all of the setups are shown in detail in Table
\ref{tab:kine}.  The data presented in this work
 were taken during two run periods in 2003.
The first period was in April and measured the mostly non-parallel
 cross sections
for $Q^2=0.060$ and 0.200 \gevc in addition to an
extension of the $Q^2=0.127$ \gevc data set.  The October period was
used to measure the $W$ scans at $Q^2=0.060$ and 0.200 \gevc and the low
$W$ background terms.
In addition to the $Q^2=0.060,0.127$, and 0.200 \gevc measurements, 
some cross check
measurements were made at $Q^2=0.127$ \gevc which overlapped with
existing data from Bates.

\begin{table}
\caption{\label{tab:kine}Kinematic values for $W$, $Q^2$, proton
  center-of-mass polar angle $\theta_{pq}^*$, proton azimuthal angle
  $\phi_{pq}^*$ and the initial electron beam energy  $E_{beam}$.  
 See Tables \ref{table:results:lowQ:XS} and \ref{table:results:Q20:XS} for
  detailed settings for the $W$ scans.
}
\begin{ruledtabular}
\begin{tabular}{ccccc}
$Q^2$ & $W$ & $\theta_{pq}^*$ & $\phi_{pq}^*$ & $E_{beam}$ \\
$[$\gevc$]$ & $[$MeV$]$ & $[^\circ]$ & $[^\circ]$ & $[$MeV$]$\\
\hline
0.060  & 1221  & ---    & $\vec{q}$ & 795\\
0.060  & 1221  & 24 & 0.0,90,180 &795 \\
0.060  & 1221  & 30 & 29 & 795\\
0.060  & 1221  & 37 & 134,180 &795\\
\hline
0.200  & 1221  & ---    & $\vec{q}$ &855\\
0.200  & 1221  & 33 & 0.0,90,180 &855\\
0.200  & 1221  & 57 & 38, 142,180 &855\\
\hline
0.060  & 1125-1300  & ---    & $\vec{q}$ &705\\
\hline
0.200  & 1125-1275  & ---    & $\vec{q}$ &855\\
\hline
0.300  & 1205  & --- & $\vec{q}$ &855\\
\hline
0.060  & 1155  & 26    & 0,180 &855\\
\hline
0.127 & 1140 & 59 & 45,135 &855\\
\hline
0.127  & 1221  & 30,43,63    & 90,135,150 & 855\\
\hline
0.127  & 1212,1232  & ---    & $\vec{q}$ & 855  \\
0.127  & 1232  & 28    & 0,180 &855\\
\end{tabular}
\end{ruledtabular}
\end{table}

\section{Data Analysis}

\subsection{Phase Space Acceptance and Simulation}

\begin{figure}
\includegraphics[angle=0,width=0.95\columnwidth]{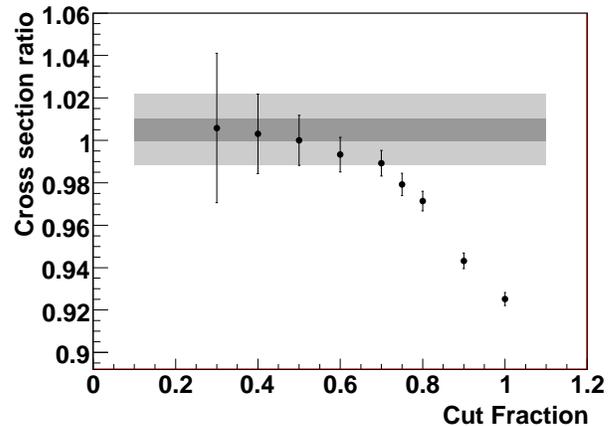}
\caption{\label{fig:analysis:scans:otherQ06}Variation in the cross
  section due to changes in cut size for $Q^2=0.060$
  \gevc, $\theta_{pq}^*=24^\circ$, $\phi_{pq}^*=0^\circ$.  
The abscissa shows the fractional phase space selection width for all the variables
 mentioned in the text and the ordinate
shows the cross section normalized to the cross section result with the 0.50
fractional phase space selection width.  
The bars are
  centered on the cut corrected result.  The dark bar is the
  statistical uncertainty and the light bar shows the total uncertainty with the
  systematic uncertainty added in quadrature.
}
\end{figure}

The phase space acceptance in the spectrometers is in a
multi-dimensional space and has a complex shape (see
Fig.~\ref{fig:acc_overlap}).  
One challenge was
defining the phase space acceptance in a similar manner across all the
kinematic settings with the phase space varying by a large amount
across the spectrometer acceptance.  One solution is to have a very small
acceptance which will limit the variations but also limit the
statistics.  Too large of a phase space acceptance leads to large
systematic errors as the variation in phase space 
is too large for the simulation to
reliably calculate.  However, a compromise can be found by settling in
a region where the combination of the statistical and systematic
errors is a minimum.  This is illustrated in
Fig.~\ref{fig:analysis:scans:otherQ06} where the normalized cross
section is plotted against the size of the kinematic cut.  More
details of the calculation are presented below but the effect of too
small of a phase space acceptance (large statistical errors on the left side of
Fig.~\ref{fig:analysis:scans:otherQ06}) can be seen as well as the
effect of too large of a phase space acceptance (large systematic
errors seen as deviations from
the central value on the right side of
Fig.~\ref{fig:analysis:scans:otherQ06}).  

To ensure uniformity of the phase space selection across the varying
kinematics, a unique solution was found which non-arbitrarily defined
the edges of the acceptance.  The maximum allowed phase space region was
found by locating the half-maximum points in the distributions of the
variables upon which the cross section depends: $W$, $Q^2$,
$\theta_{pq}^{*}$, $\phi_{pq}^{*}$.  Symmetry around the central
kinematics was also enforced so that neither side of the phase space
was weighted too heavily.  Once the maximum acceptance regions were defined, then
fractional widths of those regions were used to study the behavior of the
extracted cross section.  Those studies, detailed below, were then
used to define the final phase space region used for extracting the cross sections.

Simul++~\cite{COLA}  is the software which was employed to calculate
the multi-dimensional phase space.  Simul++ also simulates
the collimators inside the spectrometers as part of calculating the
phase space.  After the subtraction of background events
(see next subsection), the spectrometer acceptance was limited in software to
the central region of the spectrometers to keep edge effects out of
the analysis.   The details are
in~\cite{stave_thesis}.  In addition to precise spectrometer
properties and collimators, Simul++ also calculates energy loss and the
radiative corrections in the same way as for the data.   Each simulated
event contains the proper weighting for radiative corrections, the
virtual photon flux $\Gamma$ and the lab to center-of-mass Jacobian.  The simulated events undergo kinematic selection processes
identical to those used on the data and can then be used to
determine the phase space and, finally, a cross section.

\begin{figure*}
\includegraphics[angle=0,width=1.5\columnwidth]{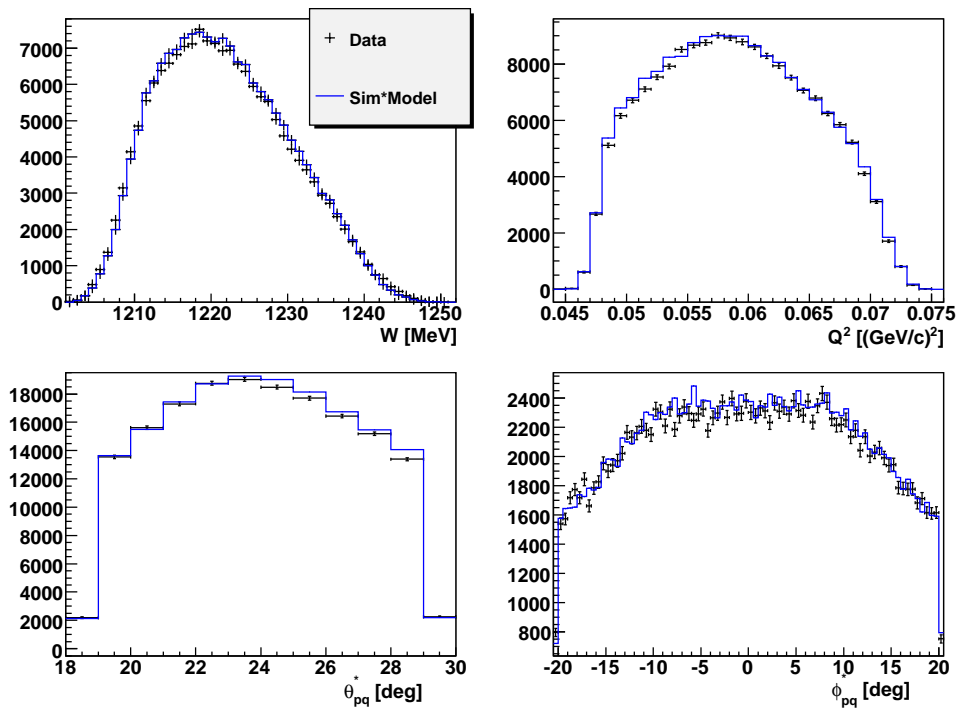}
\caption{\label{fig:data_sim}(Color online) Comparison of the relative shapes in
  phase space for the data and the
  simulation weighted by MAID 2003~\cite{maid1} for the primary variables 
  used in the cut.  The results are for an in-plane, forward angle setup at
  $Q^2=0.060$ \gevc and show a good level of agreement.}
\end{figure*}

\begin{figure}
\includegraphics[angle=0,width=0.95\columnwidth]{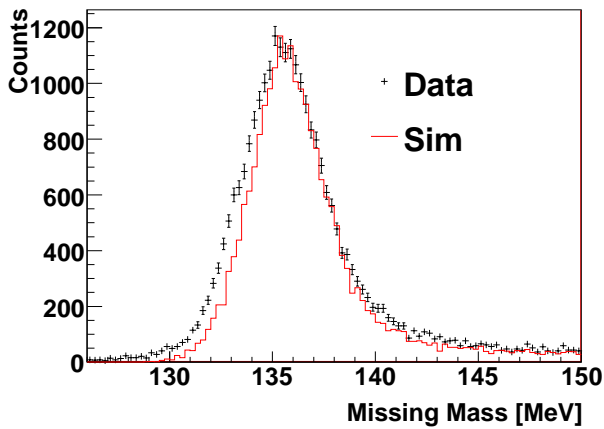}
\caption{\label{fig:mm_data_sim}(Color online) Comparison of the missing
  mass from the data (black crosses) and the simulation (solid line)
  for the $Q^2=0.060$ \gevc, $W=1221$ MeV, $\theta_{pq}^*=36^\circ$,
  $\phi_{pq}^*=180^\circ$ setting. }
\end{figure}

Figure \ref{fig:data_sim} shows
 a comparison of the data for an in-plane, forward setup with the
results of Simul++ weighted by the MAID 2003 phenomenological model~\cite{maid1} cross section and plotted
 against the four physics variables upon which the cross section depends:
 $W$, $Q^2$, $\theta_{pq}^{*}$, $\phi_{pq}^{*}$.  As is clear in the
 figure, there is very good agreement for
 all the variables across the acceptance.  A fifth variable,
 $z$, was also examined closely because it affects the size of the spectrometer acceptance.  $z$ is the
 vertex position determined by Spectrometer B which 
has better vertex resolution than Spectrometer A.  
The real
 edges of the $z$ distribution are not as sharp as in the simulation,
but extensive studies 
showed that avoiding those regions in $z$ 
yielded reliable cross section results.
Other setups have similarly good agreement between data and simulation. 
In addition to good agreement in the previously listed variables,
there is also acceptable agreement on the shape and location of the missing
mass peak as shown in Fig.~\ref{fig:mm_data_sim}.  The differences
between simulation and data for the missing mass do not cause
appreciable uncertainties and the level of agreement is sufficient for
this analysis.

 To investigate the effects of different sized phase space acceptance
 regions on the extracted cross section, several
types of studies were performed.  In all of them, the maximum phase
space selection width was defined by the half-maximum points as
mentioned earlier.  Then, only the fractional
width of 
the phase space region relative to the maximum
width was varied.  In this manner, the phase space selection was
consistent across the many kinematic settings.

Fine scans were made for each
kinematic setting by simultaneously 
varying the fractional width of the phase space
selection of all the physics variables 
($W$, $Q^2$, $\theta_{pq}^*$, $\phi_{pq}^*$) and the vertex position, $z$.
(The cross section does not depend on the vertex position but the
shape of the vertex distribution did change from setup to setup
requiring a similar definition of the cut.)
Figure~\ref{fig:analysis:scans:otherQ06} 
shows the scan for a
non-parallel, forward angle, $Q^2=0.060$ \gevc setting.
 The abscissa shows the fractional phase space selection width for all the variables
 mentioned and the ordinate
shows the cross section normalized to the cross section result with the 0.50
fractional phase space selection width.  
The variation of the cross section ratio with changing cut fractions
seen in Fig.~\ref{fig:analysis:scans:otherQ06} was representative of the variation seen in the other kinematic
settings.  As mentioned before, 
Figure~\ref{fig:analysis:scans:otherQ06} shows that the
extracted cross section gets more stable with smaller selection regions but the
statistical uncertainties necessarily get larger.  Small statistical uncertainties
are possible with larger cut fractions but then the systematic errors suffer.  What is not shown here is
that the helicity dependent cross sections have the most stable results
for a fractional phase space selection width of 0.75.  The fraction of 0.75 was then chosen 
as a compromise in order to have stable results across
all kinematics with small statistical uncertainties.  

In most
settings, like that shown in Fig.~\ref{fig:analysis:scans:otherQ06}, the
cross section for a fractional phase space selection width of 0.75 is slightly lower than the 0.50 result.  To
correct for this, a phase space correction factor was determined by
averaging the results of these selection scans over all the similar
kinematic settings.   These corrections are all on the
order of 2 to 3\% with a 1\% systematic uncertainty.  (Some of the backward angle settings had
flat phase space selection width scan results and did not require a phase space correction factor.)  For a comparison
of the relative uncertainties, see the light and
dark bars in Fig.~\ref{fig:analysis:scans:otherQ06}.  The light bar shows the
statistical uncertainty for the final cross section result with a
fractional phase space selection width
of 0.75 and the dark bar shows the total 
uncertainty including the appropriate systematic uncertainties.  (Since this is a comparison of only one kinematic
setting, it is not appropriate to include any uncertainties from
quantities which vary statistically or systematically from
setup to setup like the luminosity.)  The cross section ratios with
fractional phase space selection widths of 0.50
and 0.75 agree with each other within statistical and systematic
uncertainty and are also both stable.

This phase space selection procedure was used for the analysis of the
$Q^2=0.060$ \gevc data.  Similar cuts were used in the $Q^2=0.200$
\gevc analysis and their stability was verified.  Any small
differences in the cuts lead to only small differences in the cross
sections and are not significant.

\subsection{Elimination of Background Counts}
During the pion production runs, there were two types of backgrounds:
$\pi^-/\mu^-$ background and general accidental background.  The
$\pi^-/\mu^-$ 
background was removed by making a two-dimensional selection
in missing mass versus coincidence timing space.  The $\pi^-/\mu^-$
background region was very clearly separated from $\pi^0$ events of
interest.  Their identity was confirmed with a Cherenkov counter in
Spectrometer A which was only present during the first running
period.  However, the two-dimensional selection in missing mass versus
coincidence timing space was found to be just as effective at removing
the $\pi^-/\mu^-$ background.

\begin{figure}
\includegraphics[angle=0,width=0.95\columnwidth]{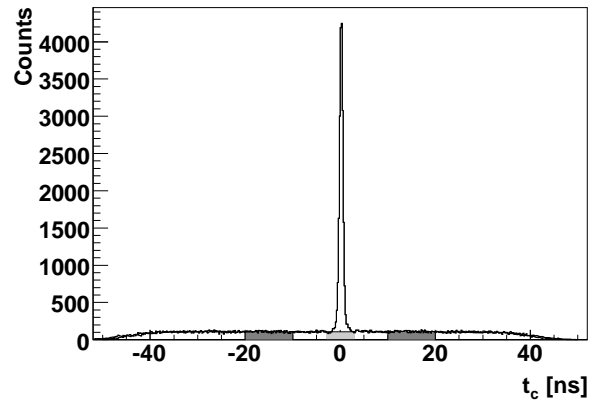}
\caption{\label{fig:bg_sub}Coincidence timing of the proton and
  electron.  
The dark gray areas indicate the  background subtraction
  region and the light gray under the peak indicates the size of the
  background.  The full-width at half-maximum of the peak is 0.9 ns.}
\end{figure}

After the $\pi^-/\mu^-$ cut, an accidental subtraction was applied
using accidentals from both sides of the coincidence peak to
determine the background counts per channel.  
Figure~\ref{fig:bg_sub} shows the coincidence peak on top of the
accidental background with the light gray region indicating the 
average background level seen in the two side regions.  
The accidental subtraction removes about 6\% to 20\% of the
events in the coincidence peak depending on the kinematics and 
is the largest of the background subtractions. 

After both the pion and accidental background subtractions, the data
consist of only coincidence events and a cross section can be
extracted which is not contaminated with background events.

\subsection{Luminosity}

The luminosity is calculated based upon the total current measured by
the F\"orster probe, a pair of toroidal coils which surround the beam
and measure the current induced by the beam \cite{sirca}.
The F\"orster probe is located in the third
stage of the microtron which can recirculate the beam up to 90 times.
Therefore, the current of the recirculated beam in the third stage can
be up to 90 times larger than the beam on target.  A measurement there
leads to a much more precise determination of the beam current.

The luminosity can then be calculated given the beam current, target
length, and target density (from pressure and temperature).  To
prevent local boiling of the hydrogen target, the electron beam is
rasterized or wobbled across the target in a rectangular pattern.  
During the April run, the beam was also placed
off-center to ensure a path to the out-of-plane Spectrometer
B that was free of obstructions.  The flat plate above the target
extended out and would have been in the path of the out-going protons if
the beam were not shifted down and to the right.  
This offset in the beam position 
decreased the effective target length by less than 1.5\% and the
effect was taken into account by the simulation.  

The normal operating pressure for the target is 2.1 bar.  With a
normal temperature of 22 K, this leads to an undercooling of 1 K.
This temperature buffer allows for 
a certain amount of local heating without the
target starting to boil.  However, both pion production run periods
experienced lower target pressure which led
to less undercooling.  Instead of 1 K undercooling, the
experiment operated closer to 0.6 K undercooling.  

\begin{figure}
\includegraphics[angle=0,width=0.95\columnwidth]{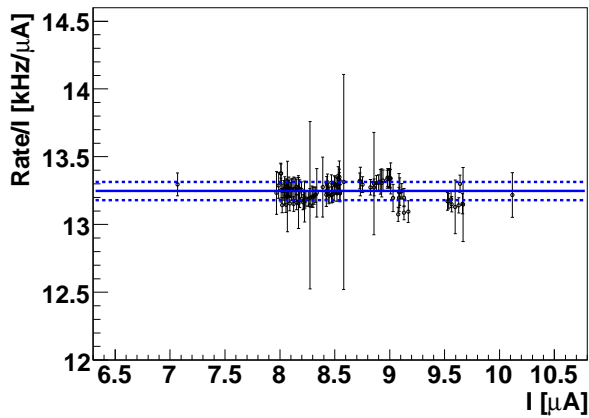}
\caption{\label{fig:lumi_corr}(Color online) The singles rate in
  Spectrometer A divided by the beam current for all the $Q^2=0.060$
  \gevc  runs in the April beam time period.  The uncertainties shown are
  statistical only.  The lines show the
  average and RMS deviation of all the data points.  Most of the data are within
  the uncertainties.  Even over a range of 3
  $\mu$A, there is no large target boiling effect. }
\end{figure}

The singles rates in Spectrometer A were used to study the effect of
the beam current on the luminosity.  (Spectrometer B was rarely in the same place from one setup to
another but Spectrometer A was returned to the same location
repeatedly.)  By plotting the singles rate in A versus beam current, any
target boiling effect should be visible.
Figure \ref{fig:lumi_corr} shows the A singles rate
divided by the beam current for all the $Q^2=0.060$ \gevc runs.  Also
plotted are the average and the RMS deviation of all of the data.  
Notice
that almost all of the points are consistent with a horizontal line which
indicates no beam current dependent luminosity change.  
Therefore, the low $Q^2$ runs were below the boiling threshold and do
not need any correction.

However, other data were taken with higher beam currents, specifically
the parallel pion production cross section comparison with Bates data
and the $Q^2=0.200$ \gevc data.  
It is possible that these runs were taken above the
boiling threshold.  To test this, the singles rates from Spectrometer
A divided by the current was plotted and a line was fit to
 the data.  During the experiment, the effect of the beam current on
 the luminosity was explicitly checked for one setting where data were
 taken at 25 $\mu$A for 2.5 hours and 12.5 $\mu$A for 5 hours.
The
results of the fit to the singles data and the beam current study
indicate a current dependent effect for beam currents above 12.5
$\mu$A.  
A luminosity correction factor and uncertainty of  $(3\pm 1)\%$ 
were adopted which are consistent with
all the available data.  More details are presented in~\cite{stave_thesis}. 

The conclusion from the luminosity studies is that the $Q^2=0.06$ \gevc data are
unaffected by beam target heating and a $(3\pm 1)\%$ correction is sufficient
to account for the effect in the remaining data.  

\subsection{Extraction at Central Kinematic Values}

The analysis procedure yields a cross section which has been averaged
over the multi-dimensional phase space while the theoretical models provide
predicted values at points in that phase space.  
To compare the averaged cross section to theory, a kinematic translation procedure
is applied to the data.  This is also known as bin centering
corrections~\cite{arrington} or transport.
The goal of the procedure is to find the correction factor which will
convert the cross section which has been averaged over phase space to
the cross section evaluated at the central kinematic values of the
phase space.  The kinematic correction factor is found by averaging
the model predictions over the same volume in phase space as the
data.  That value is then divided by the model prediction at the
center of the phase space.  The inverse of that ratio is the
correction factor.  This technique does
not rely upon the absolute size of the theory but merely requires that
the theory have the same shape throughout the same phase space as the data.
Corrections are typically 2 to 3\% indicating that the cross section
tends to vary smoothly and fairly symmetrically through the phase space.  
A small (0.5\%) systematic uncertainty is introduced with this
method which was estimated by performing the translation with several
models and taking the RMS deviation of the results.

This method of translation was tested by varying the size of the phase
space selection region
and checking for convergence to the point cross section.  Smaller cuts
lead to larger statistical uncertainties but the tests showed that the results
were stable and converged within the uncertainties. 

\subsection{Absolute Cross Section Verification}
In order to determine stability over time and the proper normalization, the
elastic reaction $p(e,e'p)$ was measured throughout the experiment.
As during the pion production runs, Spectrometer A was used to detect
electrons and Spectrometer B for protons.  The measurement 
uncertainties are dominated by the systematics estimated at
approximately 4\%.
The results are stable over time and
are consistent, within systematic uncertainties, with the 
1996 dispersion-theoretical analysis fit to the world elastic data~\cite{mergell}.  
The 2004 dispersion analysis~\cite{hammer} and other fits to the
elastic scattering data~\cite{seely,lomon2002,friedrich,simon1980}
were examined and there is only a small amount of spread between the
fits and they agree at $(98.5 \pm 1.5)\%$ of the 1996 dispersion fit.
A more recent dispersion analysis~\cite{hammer07} is slightly lower
(about 95\% of the 1996 fit) but agrees with the other fits and the
data within the systematic errors.

The conclusion from the coincidence elastic analysis is that the measured
cross sections are stable over time and agree well with previous
elastic results.  This indicates stability in the luminosity, target
density, and beam position.  It also indicates that the spectrometers can
be placed reliably (typically 0.6 mm and 0.1 mrad~\cite{blom}) and that the central momenta are well known.

\begin{figure}
\includegraphics[angle=0,width=0.95\columnwidth]{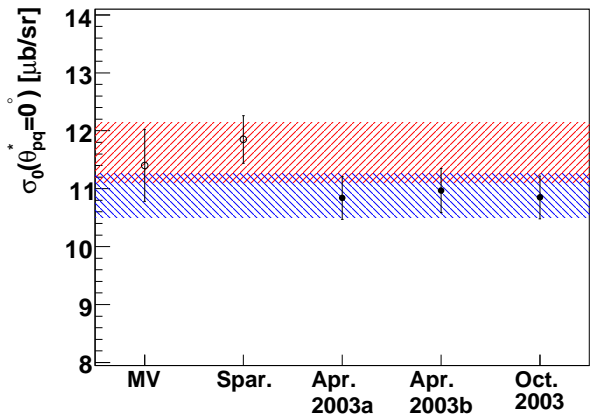}
\caption{\label{fig:analysis:par_XS_comp}(Color online) Comparison of the parallel
  cross section from previous experiments at Bates ($\circ$: Mertz-Vellidis (MV)
  $\epsilon=0.614$~\cite{mertz}
  and Sparveris (Spar.) $\epsilon=0.768$~\cite{sparveris}) and from the current
  experiment ($\bullet$) all measured at or converted to
  $\epsilon=0.707$.  The uncertainties are the statistical and
  systematic uncertainties added in quadrature.  The / lines are the average of the central values
  and the uncertainties for Bates.  The $\backslash$ lines are the values for
  Mainz.  The overlap region is easily seen.}
\end{figure}

For another check of the absolute cross sections, the parallel
pion production cross section at $W=1232$ MeV and 
$Q^2=0.127$ \gevc was taken during both run times. This measurement had been carried out previously
at Bates~\cite{mertz,mertz_thesis,vellidis,sparveris,sparveris_thesis}.  The
beam energy for each of these past experiments was slightly different
and so the $\epsilon$ factor is slightly different.  This can be
corrected for using the ratio of $\sigma_{T}$ to $\sigma_{L}$ from a
model.
Using MAID 2003, the correction factor is about 1\% and is even
smaller for other models.
Figure \ref{fig:analysis:par_XS_comp} shows all of the parallel cross
section comparisons for the previous Bates data and the current
experiment.  There is a
reasonable overlap region since the systematic uncertainties are
accounted for in the plot.  The Mainz results are stable over time
from April 2003 to October 2003.  Another item to consider is that the
variation in the Bates measurements is about 4\% and the difference
from Mainz to the lowest Bates point is about the same.  
The conclusion drawn is that the current measurement agrees with
previous measurements within the systematic uncertainties.

\subsection{Systematic Uncertainties}

As mentioned above, the uncertainties from the kinematic translation procedure can be estimated by
using various models and looking at the RMS deviation and, for most
settings, the effect is less than 0.5\%.  The one exception is a 1.3\%
effect due to worse phase space overlap in the $Q^2=0.06$ \gevc,
$\theta_{pq^*}=29.6^\circ$ setting.  This was caused by unforeseen
difficulties in placing the spectrometers.

Table \ref{table:sys_errs} summarizes the remaining uncertainties.  The
luminosity uncertainty comes from a 1\% uncertainty in the target length and
a 1\% uncertainty in the density.  Those estimates have been
conservatively added linearly.  However, the stability of the
elastic cross sections indicates that this systematic uncertainty should affect
all runs in the same way.  The detector inefficiency correction was
estimated in previous works and is quoted here~\cite{richter,sirca}.  
The dead time
correction factor was calculated using vetoed and unvetoed scalers and
is based upon counting statistics. 

The phase space cut uncertainties were found by varying the size of
the kinematic phase space cuts.
The large, in-plane angle
settings had very little difference, but for the rest of the settings,
the difference was typically 2 to 3\%.  The systematic uncertainty in phase
space cut uncertainties was estimated to be the average of the uncertainties in the
ratios of the small and large cut regions.  
The systematic uncertainty in the cut
correction is between
1.5\% and 2.4\%.  

The model uncertainty in kinematic translation has already been detailed as has the beam
current luminosity correction uncertainty.  Note that the beam current
related luminosity correction is not applied for beam currents less
than 12.5 $\mu$A and when it is applied, has a 1\% uncertainty.

To see the effect of the spectrometer angular and momentum resolution, 
the central momentum and angle settings
for the spectrometers were shifted in the simulation and the shifted simulation
results were used to extract cross sections.  Using various combinations of
the resolutions and for several, representative setups, the resolution
uncertainty was estimated at 1\%.  The spectrometer positioning uncertainties of
0.6 mm and 0.1 mrad~\cite{blom} are much smaller
than the resolution uncertainties and so do not affect the
results.  The beam position can also affect the cross section.  A study
showed that this effect is about 1\%. 

To summarize, there are several corrections applied
to the data (luminosity, phase space, kinematic translation) but they have all
been studied in detail and their contributions are all well
determined.  The total systematic uncertainties are in the range of 3 to 4\% and
agree very well with the estimates based upon comparisons with the
world elastic cross sections.


\begin{table}
\begin{ruledtabular}
\begin{tabular}{cc}
Uncertainty & Size [\%]\\
\hline
Luminosity                           & 2\\
Detector inefficiency correction     & 1 \\
Dead time correction uncertainty per setup  & $<0.5$ \\
Phase space cut uncertainty                & 1.5 - 2.4 \\
Model uncertainty              & 0.4,1.3\\
Beam current luminosity correction         &0,1\\
Momentum and angular resolution      & 1\\
Beam position                        & 1\\
\hline
Total in quadrature &                3.3 - 3.7\\
\hline
Beam polarization  &                1.2 \\
\end{tabular}
\end{ruledtabular}
\caption{\label{table:sys_errs}Summary of systematic and model uncertainties.}
\end{table}

\FloatBarrier
\section{Experimental Results}
The methods described in the previous sections were applied to the
data at $Q^2=0.060,0.127$ and $0.200$ \gevc and the results are given
here and in the
Appendix in tabular form.  There are two types of cross sections
presented.  One is the five-fold differential cross
section which is dependent on 
$W, Q^2,\theta_{pq}^*$, and $\phi_{pq}^*$ and is measured directly by the spectrometers.
The other type is the two-fold differential cross section which is
$\phi_{pq}^*$ independent and must be extracted
from the five-fold cross sections using Eq.~\ref{eq:XS}.  Both types
of cross sections
are used to aid in comparison with theory and for fitting purposes.

\subsection{Near resonance: $Q^2= 0.060, 0.200$ \gevc   \label{Q06_Q20_fit}}

\begin{figure*}
\includegraphics[angle=0,width=1.8\columnwidth]{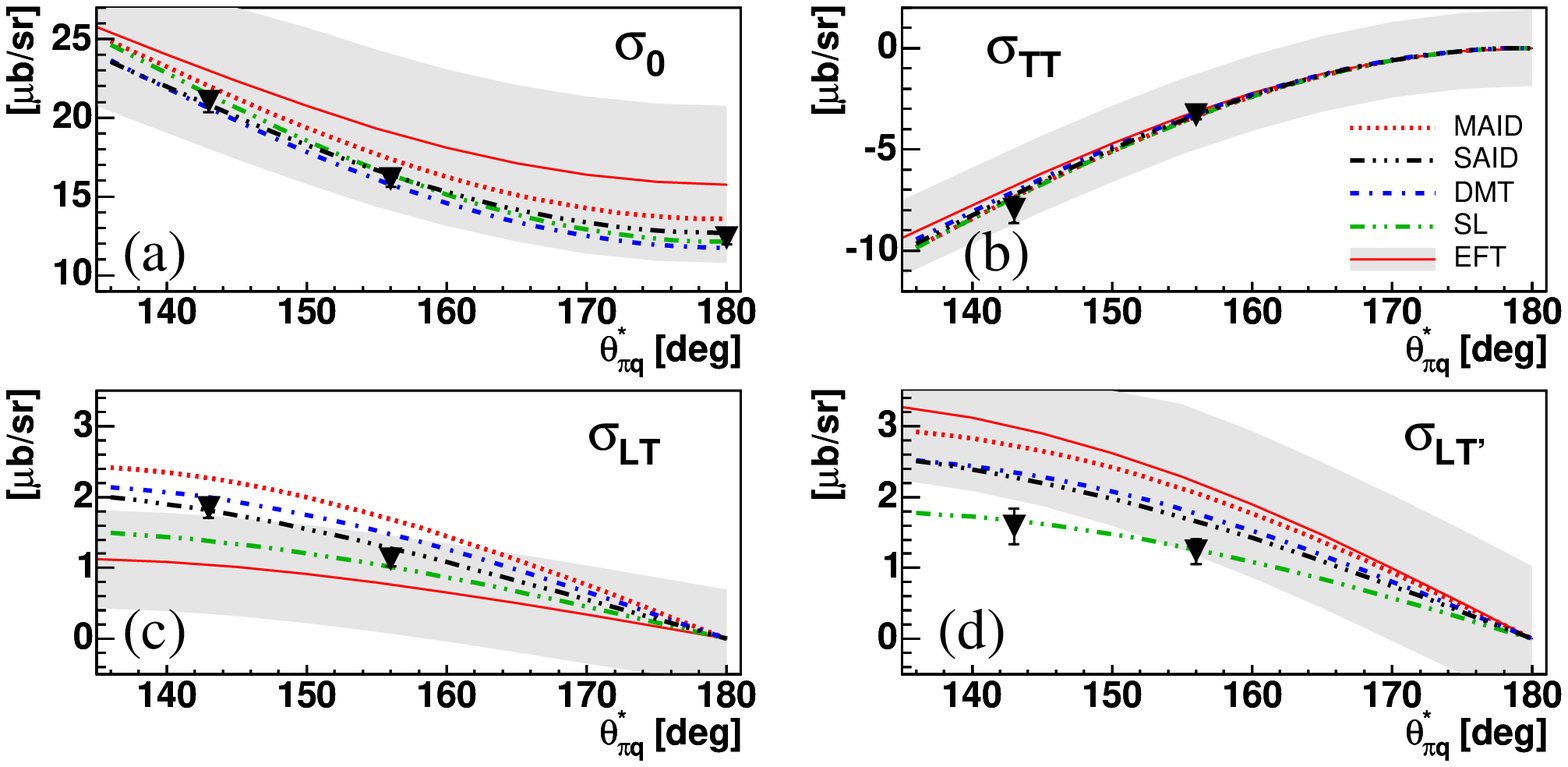}
\vspace{0.05in}
\hrule
\vspace{0.1in}
\includegraphics[angle=0,width=1.8\columnwidth]{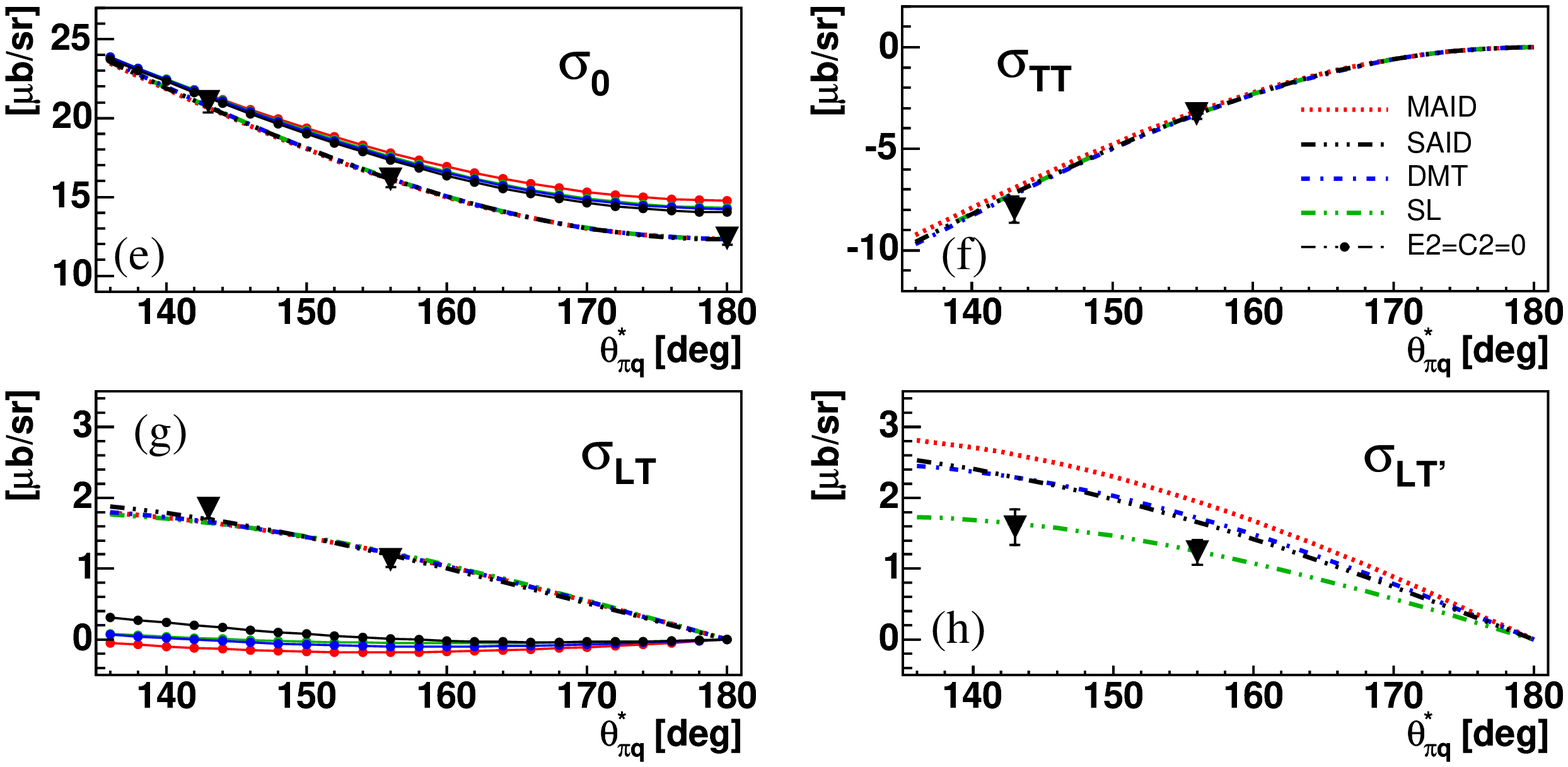}
\caption{\label{fig:Q06_XS_all}(Color online) The measured $\sigma_0=
\sigma_T +
\epsilon\sigma_L, \sigma_{TT}, \sigma_{LT}$, and $\sigma_{LT'}$
differential cross sections as a function of $\theta_{\pi q}^{*}$ at
$W=1221$ MeV and $Q^2=0.060$ (GeV/c)$^2$.
 The $\blacktriangledown$ symbols are our data
points and include the statistical and systematic uncertainties
added in quadrature. The top figures (panels a-d) show the data with the EFT
predictions~\cite{pasc} which are plotted with their estimated
uncertainties. The other curves
represent predictions from the  MAID 2003~\cite{maid1},
SL(Sato-Lee)~\cite{sato_lee}, DMT~\cite{dmt}, and SAID~\cite{said} models.
The bottom figures (panels e-h) show our data with model curves for which the three
resonant multipoles  $M_{1+}^{3/2},
E_{1+}^{3/2}, S_{1+}^{3/2}$ are fit to the data.
The lines with dots are the fitted models with the
$E_{1+}^{3/2}$ and $S_{1+}^{3/2}$ quadrupole terms set to zero and are only
plotted for the sensitive observables, $\sigma_0$ and $\sigma_{LT}$.
}
\end{figure*}

\begin{figure*}
\includegraphics[angle=0,width=1.8\columnwidth]{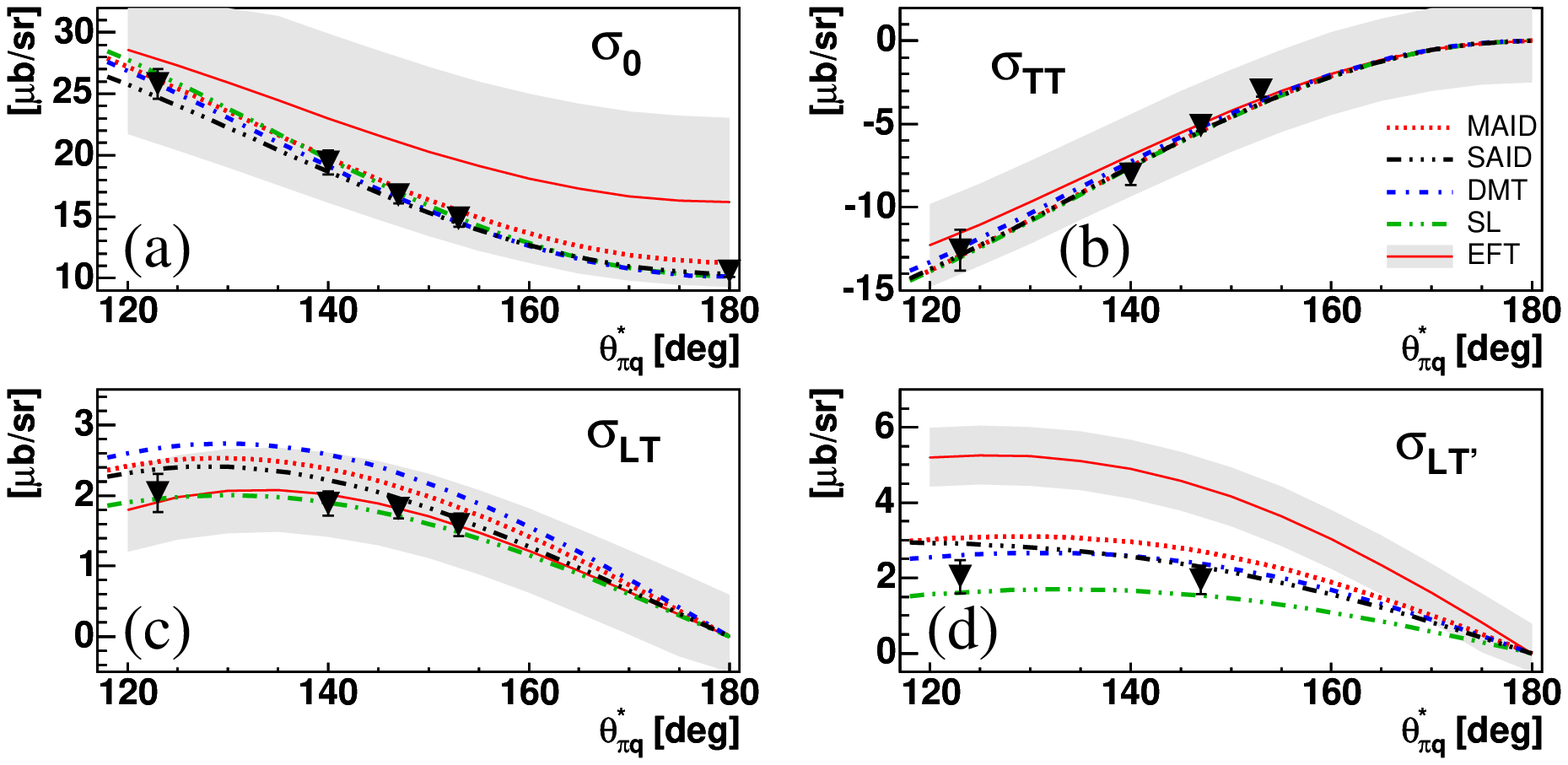}
\vspace{0.05in}
\hrule
\vspace{0.1in}
\includegraphics[angle=0,width=1.8\columnwidth]{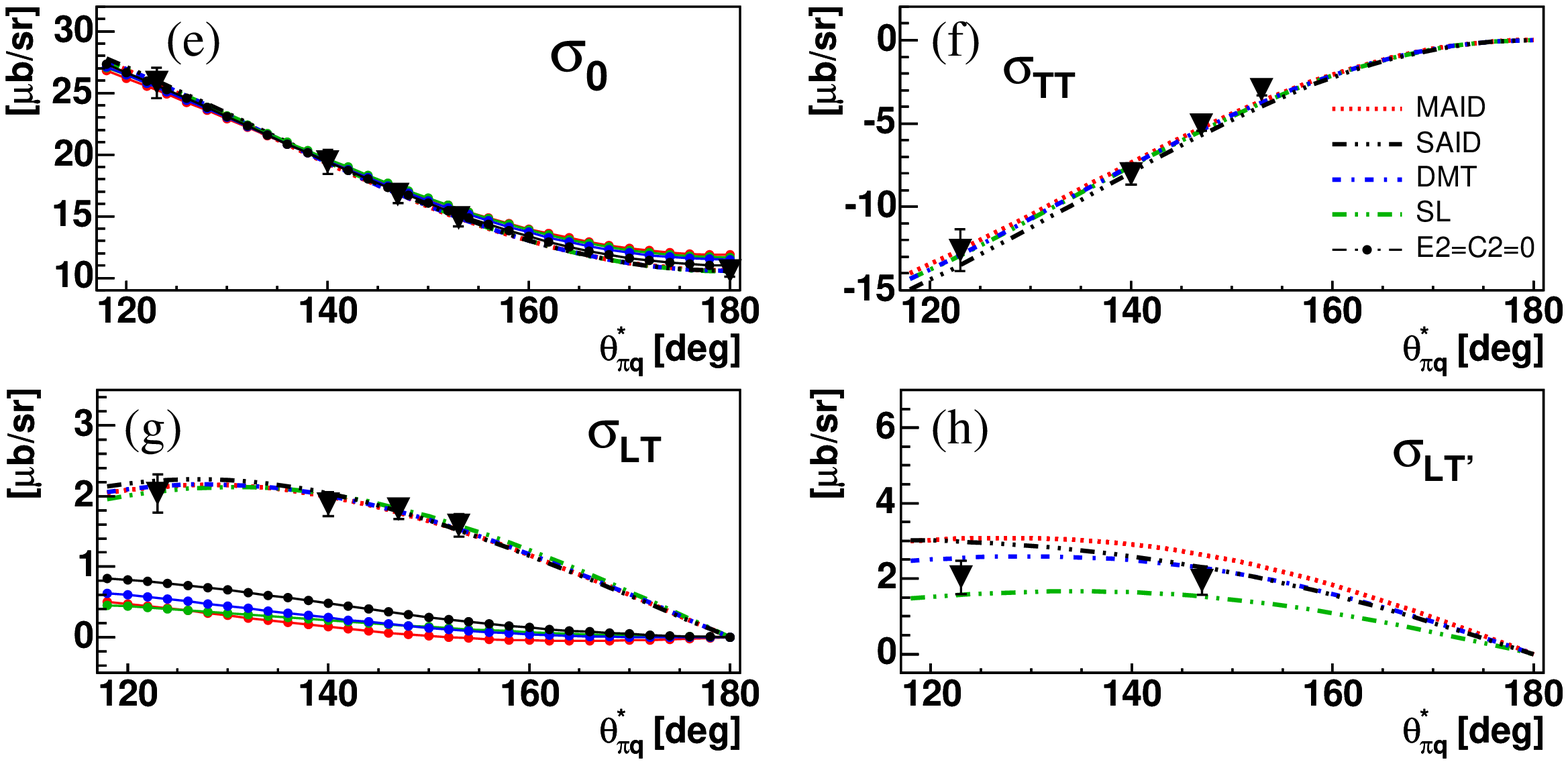}
\caption{\label{fig:Q20_XS_all}(Color online) The measured $\sigma_0=
\sigma_T +
\epsilon\sigma_L, \sigma_{TT}, \sigma_{LT}$, and $\sigma_{LT'}$
differential cross sections as a function of $\theta_{\pi q}^{*}$ at
$W=1221$ MeV and $Q^2=0.200$ (GeV/c)$^2$.
 The $\blacktriangledown$ symbols are our data
points and include the statistical and systematic uncertainties
added in quadrature. The top figures (panels a-d) show the data with the EFT
predictions~\cite{pasc} which are plotted with their estimated
uncertainties. The other curves
represent predictions from the  MAID 2003~\cite{maid1},
SL(Sato-Lee)~\cite{sato_lee}, DMT~\cite{dmt}, and SAID~\cite{said} models.
The bottom figures (panels e-h) show our data with model curves for which the three
resonant multipoles  $M_{1+}^{3/2},
E_{1+}^{3/2}, S_{1+}^{3/2}$ are fit to the data.
The lines with dots are the fitted models with the
$E_{1+}^{3/2}$ and $S_{1+}^{3/2}$ quadrupole terms set to zero and are only
plotted for the sensitive observables, $\sigma_0$ and $\sigma_{LT}$.
}
\end{figure*}

       The extracted partial cross sections $\sigma_0$, $\sigma_{TT}$,
$\sigma_{LT}$, and $\sigma_{LT'}$ versus $\theta_{\pi q}^*$ for $W =
1221$ MeV, $Q^2 =0.060$ and $W=1232$ MeV, $Q^2=0.200$ \gevc are plotted in
Figs. \ref{fig:Q06_XS_all} and \ref{fig:Q20_XS_all} respectively.
These data are compared with the chiral effective field theory
calculations (EFT) which have a few low energy parameters and then
rely on theory to arrive at results.  These calculations have 
relatively large estimated uncertainty bands due to
the neglect of higher order terms. Within these uncertainties the agreement
with experiment is good.  While these calculations and their
uncertainty estimates are a great contribution to the field,
conclusions cannot be drawn unless there are further improvements.
The precision of the data is such that more precise theory is
required.  The inclusion of even higher order terms appears to be necessary.

The top sections of Figures \ref{fig:Q06_XS_all} and \ref{fig:Q20_XS_all} also show the
predictions of four model calculations. The Sato-Lee (SL)
\cite{sato_lee} and
Dubna-Mainz-Taipei (DMT) \cite{dmt} models contain
explicit pion cloud contributions while the MAID \cite{maid1} and SAID
\cite{said} calculations are
primarily phenomenological.  These models have been adjusted by their
authors to agree with our previous data
\cite{warren,mertz,kunz,sparveris}.  For  $Q^2=0.060$ \gevcp,  all models
agree with the data for $\sigma_{TT}$. For $\sigma_0$ only MAID is 
not in agreement with the data. However, for $\sigma_{LT}$ the
dispersion between the models and data is greater showing that they
have not been adjusted to agree with $S_{1+}$. For $Q^2=0.200$ \gevcp, the
agreement between models and experiment for $S_{1+}$ is even less satisfactory.
One item that this indicates is that the $Q^2$ dependence of $S_{1+}$
is not correct in DMT and MAID since both models agree well with
$S_{1+}$ data at $Q^2=0.127$ \gevc shown in Ref.~\cite{sparveris}.
For
$\sigma_{LT'}$, only the SL model agrees with the data at both $Q^2$
values.  All of these disagreements show the importance of performing
measurements at low $Q^2$.

The extraction of the three resonant $\gamma^* + p \rightarrow \Delta$
amplitudes $M_{1+}$, $E_{1+}$ and $S_{1+}$ was accomplished by
adjusting these amplitudes in the four phenomenological models
described above.  Following the practice of
Refs.~\cite{mertz,sparveris,stave,sparveris_Q20} the model dependent
extraction from successful phenomenological reaction models allows for
a reliable extraction of the resonant amplitudes.  The model uncertainty is
estimated by the spread of the derived values using the various model
amplitudes \cite{cnp,stave_etal,stiliaris_cnp}. 

The fitting procedure used in this analysis is described in detail in~\cite{stave_thesis}.  Briefly, the procedure takes all the background
multipoles up to $L=5$ from a model and varies the amplitude of the
resonant, isospin 3/2 multipoles ($M_{1+}^{3/2}$, $E_{1+}^{3/2}$ and
$S_{1+}^{3/2}$) to attain a best fit to data at one value of $W$ and
$Q^2$.  By performing the fit in this manner, there is not the usual
truncation of the fit past p waves.  However, there is a model
dependence since the various models differ
in the sizes of background terms.

The fitting of the data started with the 
helicity independent results, the three
$\theta_{pq}^*$ angles with the $\phi_{pq}^*$ dependence.  Those seven
five-fold differential cross section results were fit using the three resonant 
parameter fit with the four models.  All the fits had $\chi^2$ per degree
of freedom near one indicating good fits.  Correlations between the
fitting parameters were taken into account in the uncertainties estimated by
the fitting routine~\cite{stave_thesis,minuit}.
Figures \ref{fig:Q06_XS_all} and \ref{fig:Q20_XS_all} show the
data and the different fitted models.  
Despite different background terms, the four model fits
converged.  It is impressive that the four model curves almost fall on
top of each other when the three resonant $\gamma^*p\rightarrow\Delta$
amplitudes ($M_{1+}^{3/2},E_{1+}^{3/2},S_{1+}^{3/2}$) are varied to
fit the data as shown in the lower panel of Figs. \ref{fig:Q06_XS_all}
and \ref{fig:Q20_XS_all}.  In addition, the lower panels show
the ``spherical'' calculated curves when the  resonant quadrupole
amplitudes ($E_{1+}^{3/2}$ in $\sigma_{0}$ and $S_{1+}^{3/2}$ in
$\sigma_{LT}$) are set equal to zero. The difference between the
spherical and full curves shows the sensitivity of these cross sections
to the quadrupole amplitudes and demonstrates the basis of the
present measurement. The small spread in the spherical curves
indicates their sensitivity to the model dependence of the background
amplitudes.

\begin{figure*}
\includegraphics[angle=0,width=1.95\columnwidth]{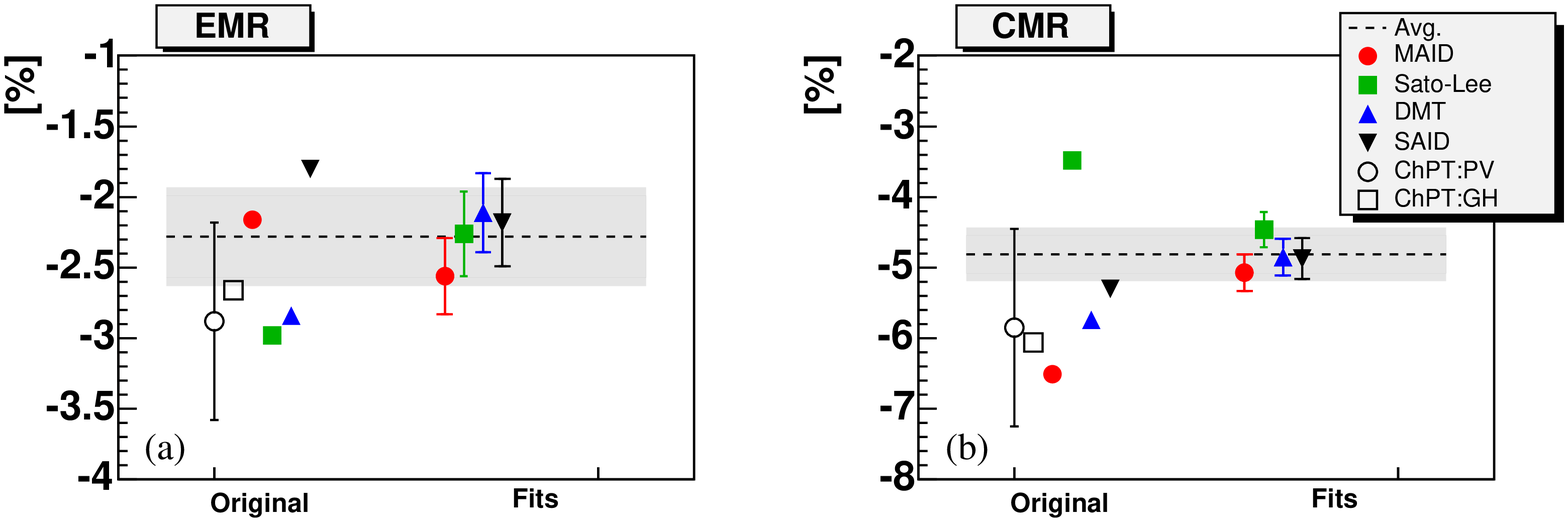}
\hrule
\includegraphics[angle=0,width=1.95\columnwidth]{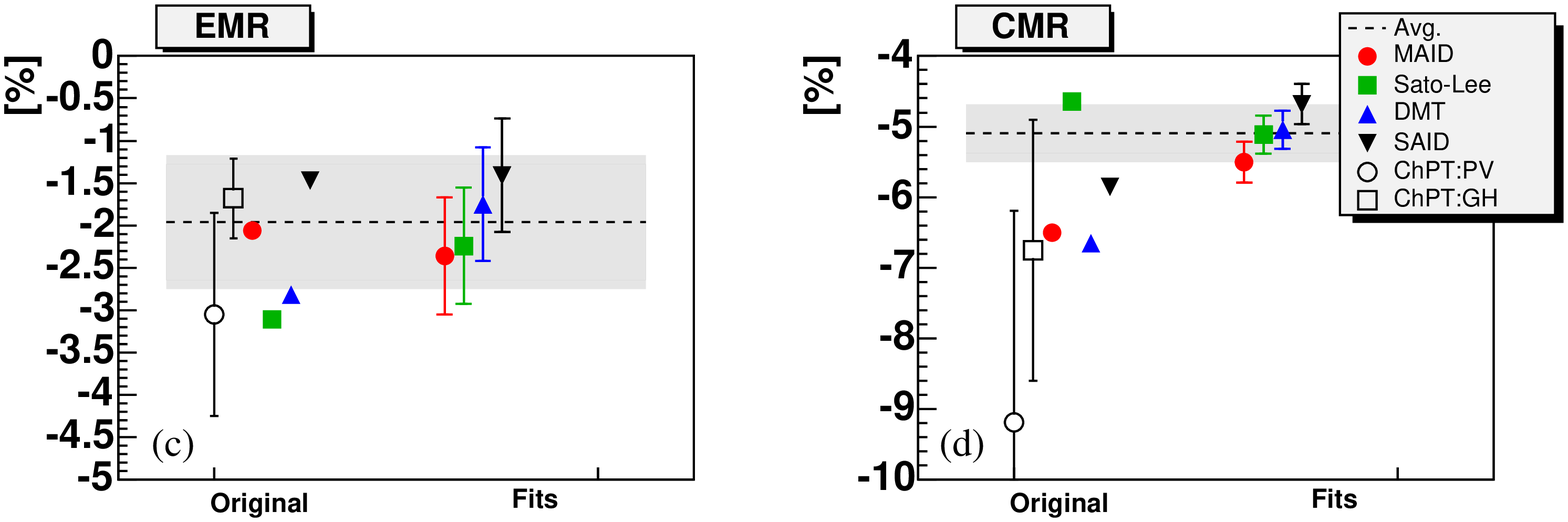}
\caption{\label{fig:results:lowQ:EMRCMR_conv}(Color online) Example of the convergence of the
  EMR and CMR values with fitting for $W=1232$ MeV, $Q^2=0.06$ \gevc
  (panels a-b) and  $W=1232$ MeV, $Q^2=0.20$ \gevc (panels c-d).  The uncertainty
  on the fits is statistical only since the systematic uncertainty is
  very small as mentioned in the text. 
  The left side of each plot shows the original model calculations
  and the right side shows the results after fitting.
The gray band is the total statistical, systematic and model
uncertainty added in quadrature.  
The models are MAID 2003~\cite{maid1,maid_web}, DMT~\cite{dmt,dmt2},
  Sato-Lee~\cite{sato_lee} and SAID~\cite{said}.  The chiral effective
  field theory predictions of Pascalutsa and Vanderhaeghen (PV)~\cite{pasc,pasc2} and Gail and Hemmert (GH)~\cite{gail_hemmert} are included.
}
\end{figure*}

Figure
\ref{fig:results:lowQ:EMRCMR_conv} shows the model convergence for the
EMR and CMR at $Q^2=0.060$ \gevc in another way.  
The convergence in $M_{1+}$ was not as significant but
the values for $M_{1+}$ have been modified by the various
model authors in order to fit previous data.  

It is interesting
that good fits were achieved in the resonant region despite the
differing model backgrounds.  
The $M_{1+}$ term is dominant but the background multipoles
are of a similar size to the resonant multipoles.  The
reason that the fitting routine is able to be rather insensitive to
the backgrounds is due, mostly,  to their having a different phase.  
Near resonance, the $I=3/2$ resonant multipoles are mostly imaginary
due to the Fermi-Watson theorem~\cite{drechsel_tiator}.  
The $E_{0+}$ and $S_{0+}$ are mostly imaginary while others are primarily real.
Since the $M_{1+}$ amplitude near resonance is almost pure imaginary, the
interference with mostly real amplitudes is very small.  In addition, 
the $E_{0+}$ multipole does not differ very much from model to
model so while it has a large effect, it does not affect the resonant
fits.  The fitting procedure is also insensitive
to the background amplitudes partly because of their angular dependence.
The primary
contributors to the cross section near resonance are the resonant
$M_{1+}$, $E_{1+}$ and $S_{1+}$ and the background $E_{0+}$ and $S_{0+}$.  
The multipole contributions to the cross section have
different angular shapes which the fitter can use to separate the
components.


As mentioned before, 
in Figures \ref{fig:Q06_XS_all} and \ref{fig:Q20_XS_all}, the
$\sigma_{LT'}$ results are only close for the Sato-Lee model but then
those cross sections were not included in this fit.  The
$\sigma_{LT'}$ cross section is sensitive primarily to the background
amplitudes and a resonant fit is not expected to improve the
agreement.  In fact, the fit results were the same, within the uncertainties,
whether or not the $\sigma_{LT'}$ data were included.


Table \ref{table:fit:bothQ} shows model and chiral EFT predictions
along with fitted results for the models and the averages of those
models at both $Q^2$ values.  The table also contains three different types of uncertainties:
statistical (used when fitting the data), systematic, and model.  The
systematic uncertainties are calculated by scaling all of the cross sections
to the minimum and maximum allowed by the uncertainties and refitting.  The range
of the refit values then gives the systematic uncertainty.   The
systematic uncertainty for the EMR and CMR mostly cancelled because
the quantities are ratios of multipoles and so are
supressed in Table~\ref{table:fit:bothQ}.  
However, since $M_{1+}^{3/2}$ is not a ratio,
the systematic uncertainties remained.  Following our
previous work \cite{sparveris,cnp,stave_etal,stiliaris_cnp}, the model
uncertainties were found by taking the root mean square deviation of the
results using the four models.  We believe that this is reasonable
since the chosen models represent state-of-the-art calculations and
also a variety of different approaches.  The final statistical and systematic
uncertainties are the average over the four models.   The model uncertainties and
experimental uncertainties are very similar in size, especially for the EMR
and CMR, as also seen at $Q^2=0.127$ \gevc in
Ref.~\cite{sparveris}.  Therefore, one can conclude that the current
experimental limit has been reached and further gains can only be
achieved after improving the models.  The effect of background
amplitudes on the resonant amplitudes was studied and determined to
have an effect approximately the same size as the model to model RMS
deviation.  This study is detailed in Refs.~\cite{stave_thesis} and~\cite{stave_etal}.

\begin{table*}
\begin{ruledtabular}
\begin{tabular}{cc|lc|lc|lc}
$Q^2$ \gevc &Model  & ~~~~~EMR (\%) &  & ~~~~~CMR (\%)&  & ~~~~~$M_{1+}^{3/2}$ ($10^{-3}/m_{\pi^+}$)& \\
            &       & ~~~Fit & Orig. & ~~~Fit & Orig. & ~~~Fit & Orig.\\
\hline
0.06 & SAID & $-2.18 \pm 0.31$ & $-1.80$ & $-4.87 \pm 0.29$ &$-5.30$  &
$40.81 \pm 0.29\pm 0.57$ & 40.72\\
     & SL & $-2.26 \pm 0.30$ & $-2.98$  & $-4.46 \pm 0.25$ & $-3.48$
     & 
     $40.20 \pm 0.27 \pm 0.56$ & 41.28 \\
     & DMT & $-2.11 \pm 0.28$ & $-2.84$  & $-4.85 \pm 0.26$ & $-5.74$  
     & $40.78 \pm 0.27 \pm 0.57$ & 40.81\\
     & MAID & $-2.56 \pm 0.27$ & $-2.16$  & $-5.07\pm 0.26$ & $-6.51$  
     & $39.51 \pm 0.26 \pm 0.57$ & 40.53\\  
\hline
     & Avg. & $-2.28 \pm 0.29 \pm 0.01 \pm 0.20$ & & $-4.81 \pm 0.27
     \pm 0.03 \pm 0.26$ & & $40.33 \pm 0.27 \pm 0.57 \pm 0.61$ & \\
\hline
     & GH   & $-2.66$ & & $-6.06$ & & 41.15 & \\
     & PV   & $-2.88 \pm 0.70$ & & $-5.85 \pm 1.40$ & & $39.75 \pm 3.87$ & \\
\hline
\hline
$Q^2$ \gevc &Model  & ~~~~~EMR (\%) &  & ~~~~~CMR (\%)&  & ~~~~~$M_{1+}^{3/2}$ ($10^{-3}/m_{\pi^+}$)& \\
            &       & ~~~Fit & Orig. & ~~~Fit & Orig. & ~~~Fit & Orig.\\
\hline
0.20 & SAID & $-1.41 \pm 0.67$ & $-1.47$ & $-4.68 \pm 0.28$ & $-5.85$  
     & $38.89 \pm 0.44 \pm 0.62$ & 39.85 \\
     & SL & $-2.24 \pm 0.69$ & $-3.11$  & $-5.11 \pm 0.27$ & $-4.64$  
     & $39.76 \pm 0.43 \pm 0.62$ & 40.48  \\
     & DMT & $-1.75 \pm 0.67$ & $-2.82$  & $-5.04 \pm 0.27$ & $-6.65$  
     & $39.84 \pm 0.43 \pm 0.62$ & 39.65 \\
     & MAID & $-2.36 \pm 0.69$ & $-2.06$  & $-5.50 \pm 0.29$ & $-6.50$  
     & $39.43 \pm 0.43 \pm 0.62$ & 39.98\\
\hline
     & Avg. & $-1.96 \pm 0.68 \pm 0.01 \pm 0.41$  & & $-5.09 \pm 0.28
     \pm 0.02 \pm 0.30$ & & $39.57 \pm 0.43 \pm 0.62 \pm 0.40$ & \\
\hline
     & GH   & $-1.68 \pm 0.47$ & & $-6.75 \pm 1.85$ & &  &\\
     & PV   & $-3.05 \pm 1.20$ & & $-9.19 \pm 3.00$ & & $38.22 \pm
     5.10$ &\\
\end{tabular}
\end{ruledtabular}
\caption{\label{table:fit:bothQ}Values of EMR, CMR, and $M_{1+}$ at $W=1232$ MeV and $Q^2=0.060$ \gevc (top) and
  $Q^2=0.200$ \gevc (bottom) for the EFT predictions and
  fitted models.  
  The uncertainties are in the order of statistical then systematic.  The
  systematic uncertainties for the individual models' EMR and CMR are suppressed because they
  are small.  For the average, the
  third number is the model uncertainty defined as the RMS deviation of the results
  from the four different models. 
The models are the three resonant parameter 
  fitted SAID~\cite{said},
  MAID~\cite{maid1}, 
  Sato-Lee(SL)~\cite{sato_lee}, and 
  DMT~\cite{dmt,dmt2} 
  models at $W=1232$ MeV (1227.3 MeV for SAID~\cite{arndt_private})
  and $Q^2=0.060$ \gevcp.  The EFT predictions are  Gail and Hemmert
  (GH)~\cite{gail_hemmert} and  Pascalutsa and Vanderhaeghen
  (PV)~\cite{pasc,pasc2} and are presented without fitting to data.
}
\end{table*}

\FloatBarrier

\subsection{Parallel cross section\label{par_XS}}

\begin{figure}
\includegraphics[angle=0,width=0.95\columnwidth]{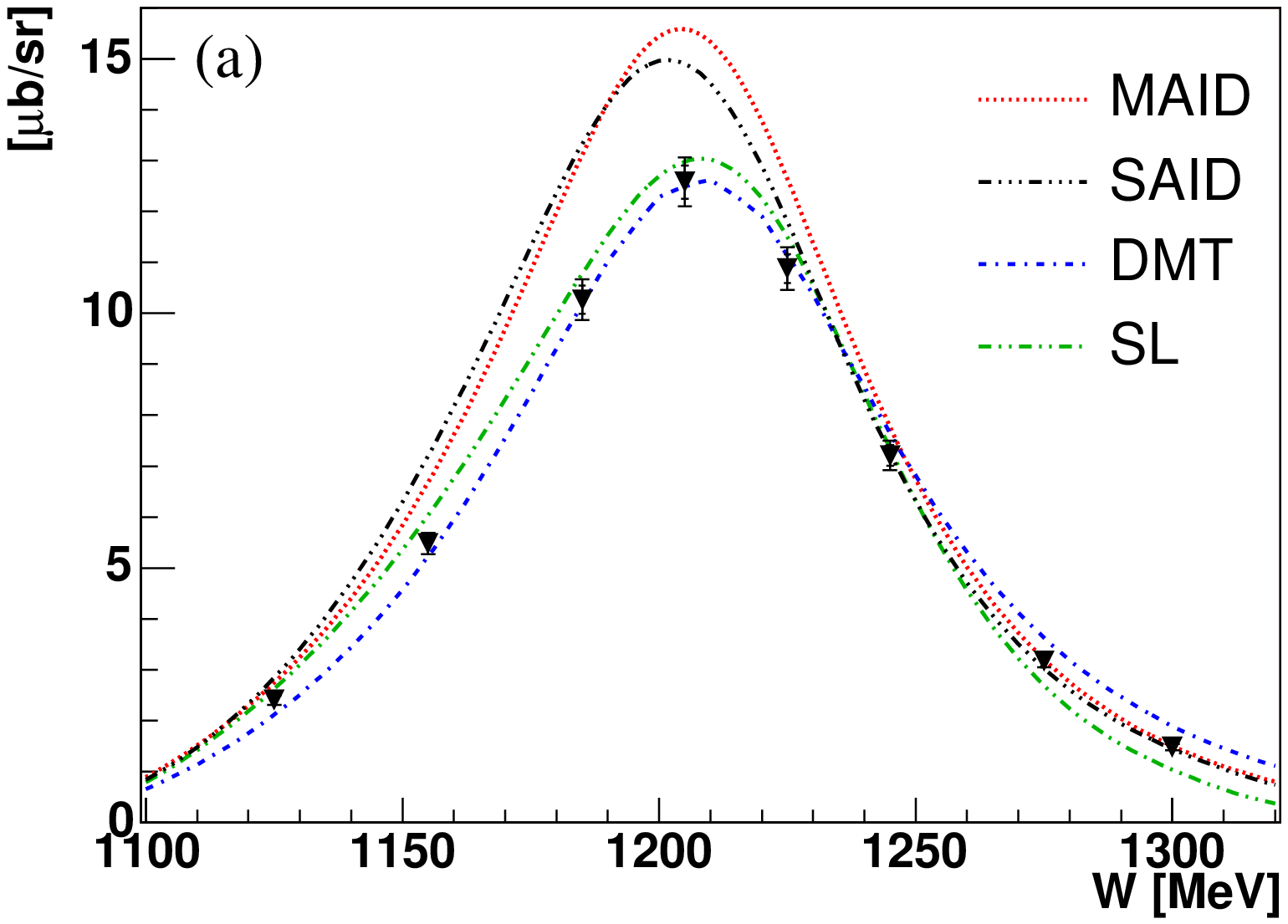}
\hrule
\vspace{0.1in}
\includegraphics[angle=0,width=0.95\columnwidth]{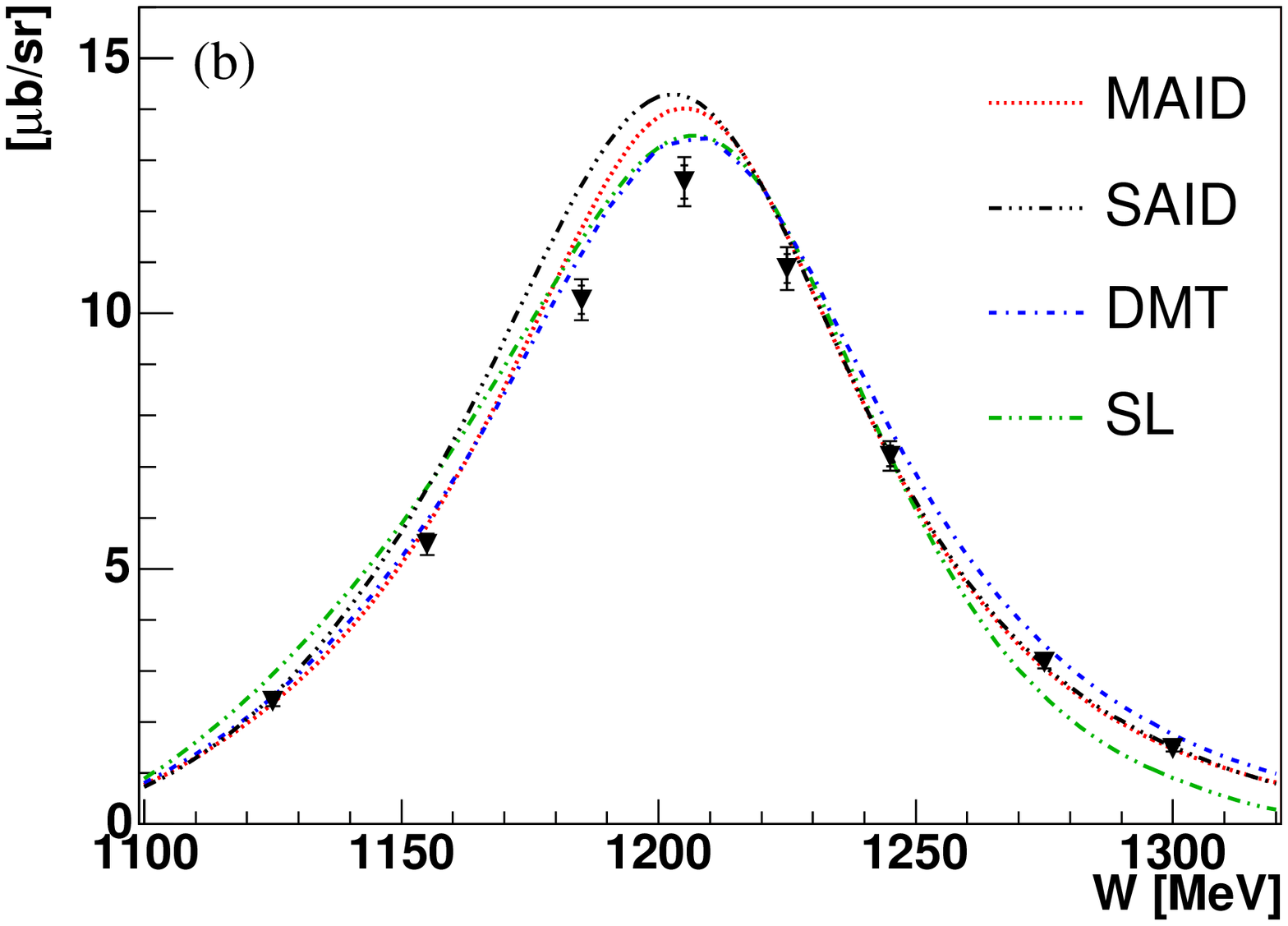}
\caption{\label{fig:Q06_W}(Color online) Parallel cross section for the
  $p(\vec{e},e'p)\pi^0$ reaction at
  $Q^2=0.060$ \gevc before (panel a) and after (panel b) three resonant
  parameter fit.  Model curves are the same as in
  Fig.~\ref{fig:Q06_XS_all}.  The smaller error bars are the
  statistical uncertainty and the larger error bars include the
  systematic uncertainty added in quadrature.} 
\end{figure}

\begin{figure}
\includegraphics[angle=0,width=0.95\columnwidth]{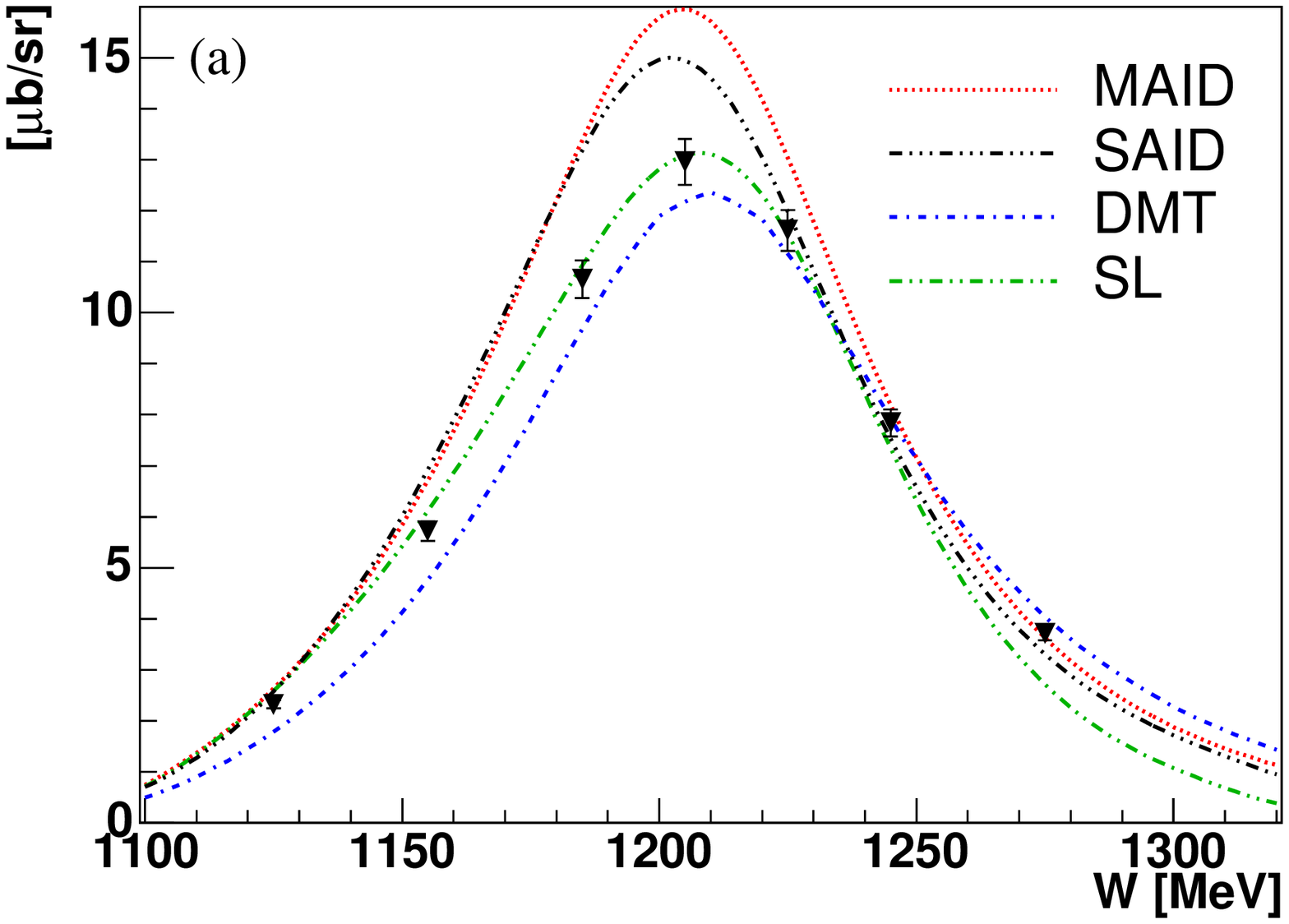}
\hrule
\includegraphics[angle=0,width=0.95\columnwidth]{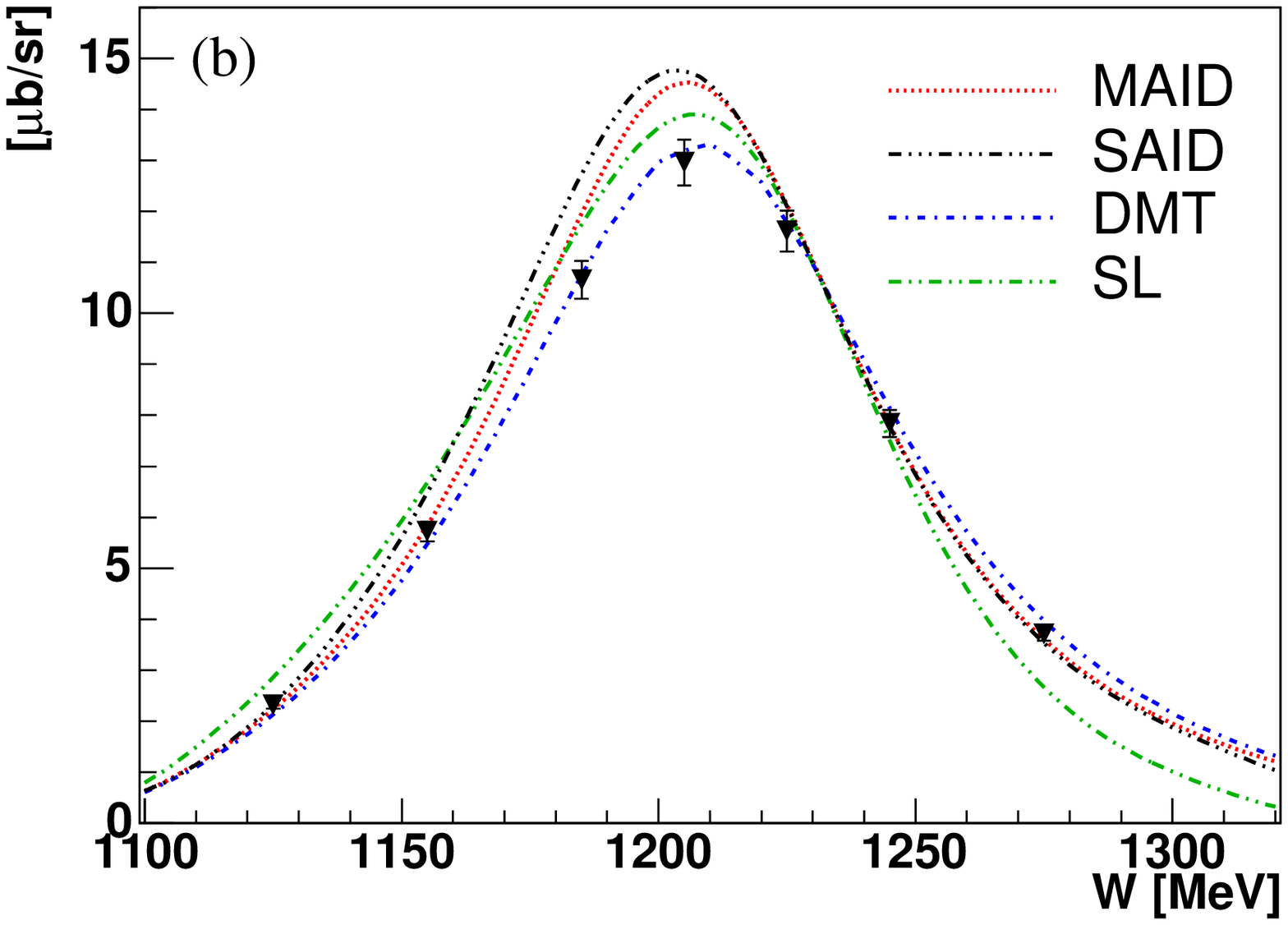}
\caption{\label{fig:Q20_W}(Color online) Parallel cross section for the
  $p(\vec{e},e'p)\pi^0$ reaction at
  $Q^2=0.200$ \gevc before (panel a) and after (panel b) fit.  Model curves
  are the same as in Fig.~\ref{fig:Q06_XS_all}. The uncertainty is the
  statistical and systematic uncertainties added in quadrature.}
\end{figure}

In Figures \ref{fig:Q06_W} and \ref{fig:Q20_W}, the parallel cross
section $W$ scans at $Q^2=0.060$ and 0.200 \gevcp, respectively, are
plotted along with corresponding model predictions.
In Panel (a) of Fig.~\ref{fig:Q06_W}, the unmodified models are shown.  In
Panel (b) of Fig.~\ref{fig:Q06_W}, the results of the three
resonant parameter fit to the previously
shown data were used.  It is important to note that in
Fig.~\ref{fig:Q06_W}, only the helicity independent, low $Q^2$
 results from Sec.~\ref{Q06_Q20_fit} have been
fit and yet the agreement with the $W$ scan data is improved significantly.
However, there is still disagreement with the data
near the tails which indicate issues with the model backgrounds.

The model curves in Fig.~\ref{fig:Q20_W} were made in the same way as
those for Fig.~\ref{fig:Q06_W}.  Again, even though the $W$ dependent
data were not included in the fit, the models converged noticeably.
There are even larger deviations at high $W$ indicating an additional $Q^2$
dependence to the model background terms which is not accounted for properly.
It is hoped that both these sets of $W$ dependent data will help to
constrain the models once the models have been improved.

\begin{figure}
\includegraphics[angle=0,width=0.95\columnwidth]{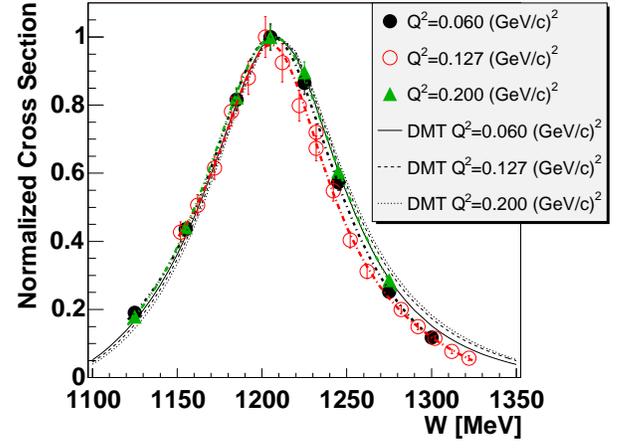}
\caption{\label{fig:par_shape}(Color online) Parallel cross section data
  (uncertainties are statistical and systematic added in quadrature) for the
  $p(\vec{e},e'p)\pi^0$ reaction from Mainz
  (this paper) and Bates~\cite{mertz,sparveris} and model
  predictions from~\cite{dmt} scaled so that the $W=1205$ MeV peak
  is at 1.0.  Note that the shape of
  the parallel cross does not vary by a large amount.  The data points
  are connected by fits (thick lines) which were a Breit-Wigner form plus a
  quadratic background.
 }
\end{figure}

One property that these data can help determine is the shape of the
parallel cross section versus $Q^2$ in the range from $Q^2=0.060$ to
0.200 \gevcp.  The four models (MAID, Sato-Lee,
SAID, and DMT) do not have a large variation of the shape of the
parallel cross section with $Q^2$ but they do differ from one another
in peak center value and width.
This was found by plotting the
model predictions versus $W$ for the $Q^2$ range of the data after
normalizing the cross sections to the value at the peak.  The same
procedure was carried out on the Mainz and Bates data and plotted with
the peak normalized DMT model in
Fig.~\ref{fig:par_shape}.  To aid in the comparison, fits to the data
were performed using a Breit-Wigner form plus a quadratic background.
The fits determined the peak to be $W=1206 \pm 1$ MeV with widths of 83 to 108
MeV with 10\% uncertainties.  
The shape of the data clearly does not change
dramatically.  There does appear to be some deviation from the
predicted shape toward high $W$ in this model.  



\begin{figure}
\includegraphics[angle=0,width=0.95\columnwidth]{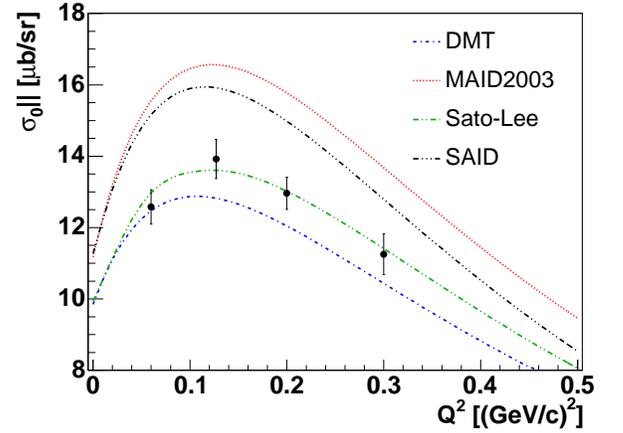}
\caption{\label{fig:Sig0_vs_Q2}(Color online) Results for $\sigma_0$ versus $Q^2$ for
$W=1205$ MeV, $\theta_{pq}^*=0^\circ$. The solid circle data were taken at
Mainz and include the statistical and systematic uncertainties added in
quadrature.  The $Q^2=0.127$ \gevc point was taken at $W=1212$ MeV and
models~\cite{maid1,sato_lee, dmt,said} were used to extrapolate to
$W=1205$ MeV.  As a result, the uncertainty is slightly larger for that point
than the others.  The plotted models have not been fit to the data. }
\end{figure}

In addition to the data taken at $Q^2=0.060,0.127,$ and $0.200$ \gevc,
one data point was taken at $Q^2=0.300$ \gevcp.  This was a parallel
cross section measurement near the peak of the cross section, $W=1205$
MeV.  As can be seen in Fig.~\ref{fig:Sig0_vs_Q2}, all of the points
taken at Mainz are consistent with the unfit Sato-Lee model but show the
same variation with $Q^2$ as all the models.   Previous data
from Bates~\cite{mertz} tend to have larger values in general as in
Fig.~\ref{fig:analysis:par_XS_comp} but are within the combined
uncertainties.  The variation with $Q^2$ is significant because the
shape is consistent with a large pion cloud contribution.  
Note that none of the models in
Fig.~\ref{fig:Sig0_vs_Q2} were fit to the data and the spread in
their predicted values is similar to the spread seen in
the Panels (a-d) of Figs.~\ref{fig:Q06_W} and \ref{fig:Q20_W} which also
show unfit models. 


\FloatBarrier

\subsection{Background Sensitive Data Below Resonance}

Background sensitive data were taken at  $Q^2=0.060$ (GeV/c)$^2$  
and at low $W$, where the $M_{1+}$ 
amplitude is less dominant, to
test the  background amplitudes in the reaction models.   
Comparing
over a wide range of $W$ is a rigorous test of the background
multipoles and the shape of $M_{1+}$.  Also, the background multipoles
are more important at low $W$ where they are relatively larger than the
resonant multipoles which are then off-resonance.   
In addition, the $M_{1+}$ term is not purely imaginary in
that region and interferences from real background amplitudes will not be
suppressed as much.  
Although the extraction of
specific background multipoles was not planned, model predictions can
be compared to the data to see whether they agree or not.  
In addition, some fitting
including background terms can provide an indication as to
 which amplitudes may be
significant.  Similar studies were performed in Ref.~\cite{sparveris}
on the Bates $Q^2=0.127$ \gevc data.  

\begin{figure}
\includegraphics[angle=0,width=1.05\columnwidth]{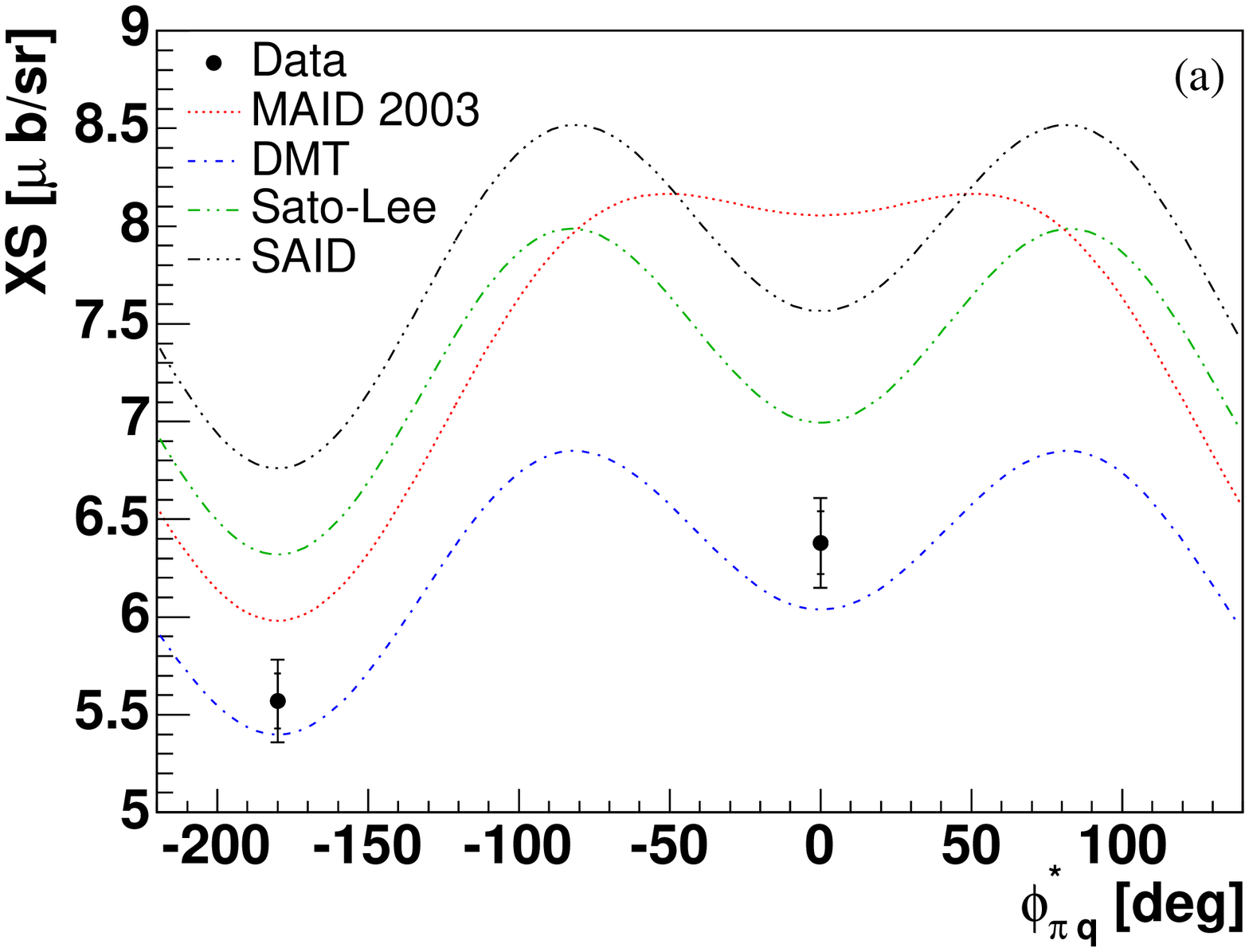}
\hrule
\vspace{0.1in}
\includegraphics[angle=0,width=1.05\columnwidth]{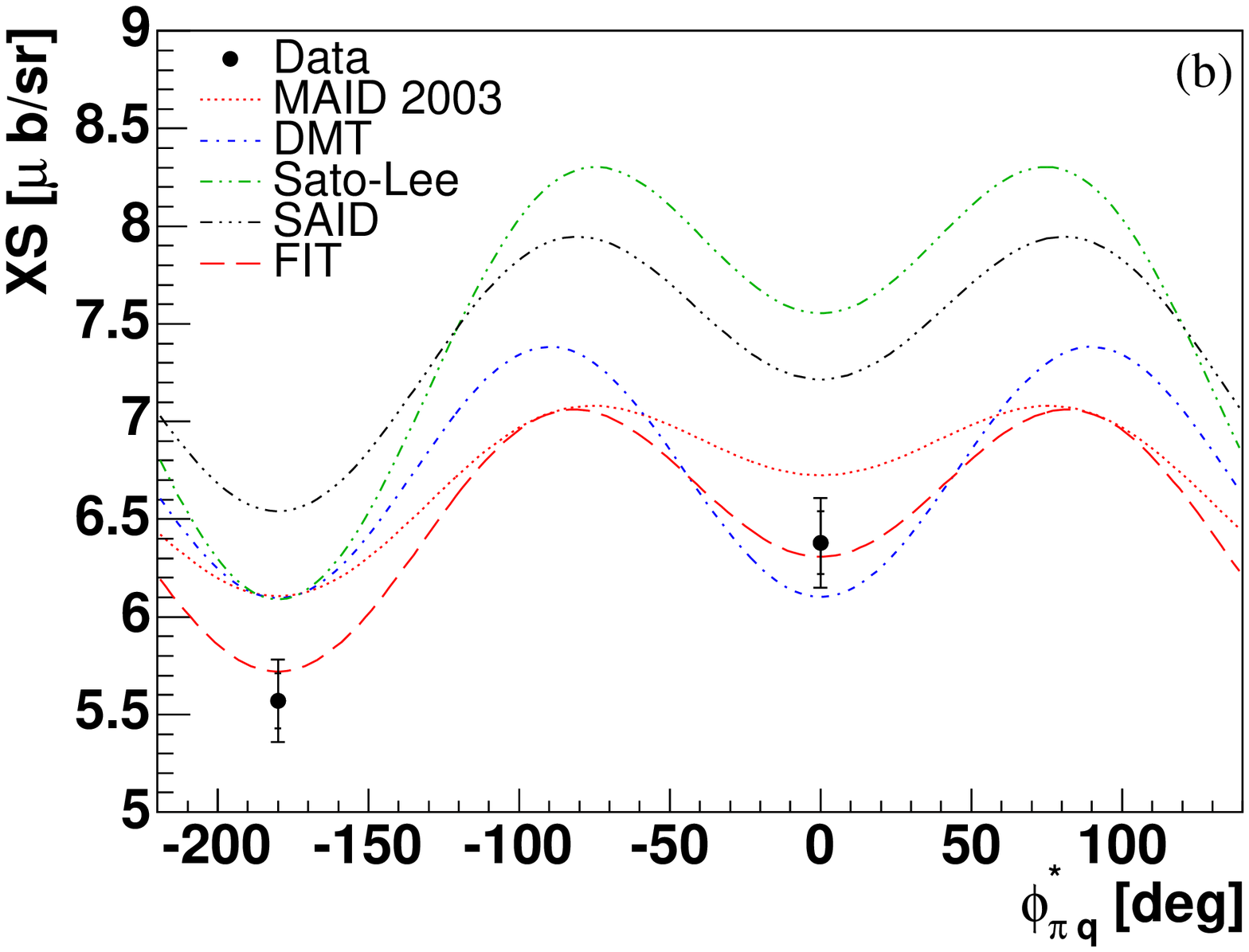}
\caption{\label{fig:BG_Q06}(Color online) Background sensitive data from Mainz at
  $W=1155$ MeV, $Q^2=0.060$
  \gevc before (panel a) and after (panel b) fit to the data near
  resonance.  The smaller error bars are the
  statistical uncertainty and the larger error bars include the
  systematic uncertainty added in quadrature.
  The curves are MAID 2003
  (dotted)~\cite{maid1} DMT (dot-dash)~\cite{dmt}, Sato-Lee
  (dot-dot-dash)~\cite{sato_lee} and SAID (dot-dot-dot-dash)~\cite{said}.
  Panel (b) also
  includes an example of one of the four parameter fits (long dash)
  including these background data.  That fit used the three resonant
  parameters along with $M_{1-}$) using the MAID 2003 model.}
\end{figure}

Figure \ref{fig:BG_Q06} shows the new low $W$
data at $Q^2=0.060$ \gevc compared with models
before and after fitting the three resonant multipoles at resonance.  
In the unfit panel (a) of Fig. \ref{fig:BG_Q06} 
only the DMT model is close to the data but does not
reproduce both points.  
The other models differ in size and shape due
to their different background amplitudes.  
The same low $W$ data points are plotted in the lower panel of
Fig. \ref{fig:BG_Q06}
along with the model fits at resonance found
in Sec. \ref{Q06_Q20_fit}.  
These data points were not
included in those fits.  The agreement with the models has not
improved significantly.  This is corroborating the known
disagreement among the models for the various background amplitudes. 

Taken together, both panels of Fig. \ref{fig:BG_Q06}
are indicating that, as expected, more than a three
parameter resonant fit is required away from the resonance region.
As an example, one of the best four parameter fits (using the
three resonant parameters along with $M_{1-}$) which include these
background data  is shown in the lower
panel of
Fig.~\ref{fig:BG_Q06}.  While that fit is satisfactory
for the two low $W$ data points, it is still in disagreement with the
background sensitive $\sigma_{LT'}$ data near resonance and for some
of the parallel cross section $W$ scan results especially at
the low
and high $W$ tails.  Because of the sensitivity of these few
points, the overall
$\chi^2$ was not improved very much.  
However, as mentioned above, since the data set is limited, 
it is not a surprise that no single
background multipole allows a good fit.  
The effect of background amplitudes from the models 
can be compared to data but the amplitudes themselves cannot be determined.

Figure \ref{fig:BG_Q06} indicates the need for more precise model
calculations and possible estimates of uncertainties.  In addition, dedicated
low $W$ experiments could help constrain the models and lead to more
refined predictions.

After comparing the data and models over a range of observables,
the DMT model has the best overall agreement with all of the low $Q^2$
data.
The fitted DMT result in Fig. \ref{fig:BG_Q06}
is the
closest to the data of all the models.  The fitted DMT results for
$\sigma_{LT'}$ in Fig. \ref{fig:Q06_XS_all} are fairly close to
the data and
no worse than MAID and SAID.
Finally, the fitted DMT results for the $W$ scan in 
Fig. \ref{fig:Q06_W} look very good overall and only disagree at
a few points.  While no model agrees perfectly with all the data, the
DMT model after the three resonant parameter fit does appear to describe the
new data the best.  

Studies using various fitting parameters indicate a path to follow for
improving the agreement between data and theory.
The problem encountered with the fitting method used in this work 
is that the parameters apparently do
not give the models enough freedom to fit the background amplitudes.  
As is clear 
in the plot in Fig.~\ref{fig:Q06_W}, the models
simply have the wrong shape.  
The next step to constrain the
background is to fit the less well determined coupling
constants and other internal model terms which affect many multipoles at once.  This
would hopefully add enough freedom to allow the models to fit the data. Despite
these quantitative problems, the background amplitudes are sufficiently
small near resonance so that the uncertainties in them do not
contribute more than the experimental uncertainties in determining the
resonant amplitudes.   

\FloatBarrier


\section{$Q^2$ variation of resonant multipoles}

\begin{figure*}
\begin{center}
\includegraphics[angle=0,height=13cm]{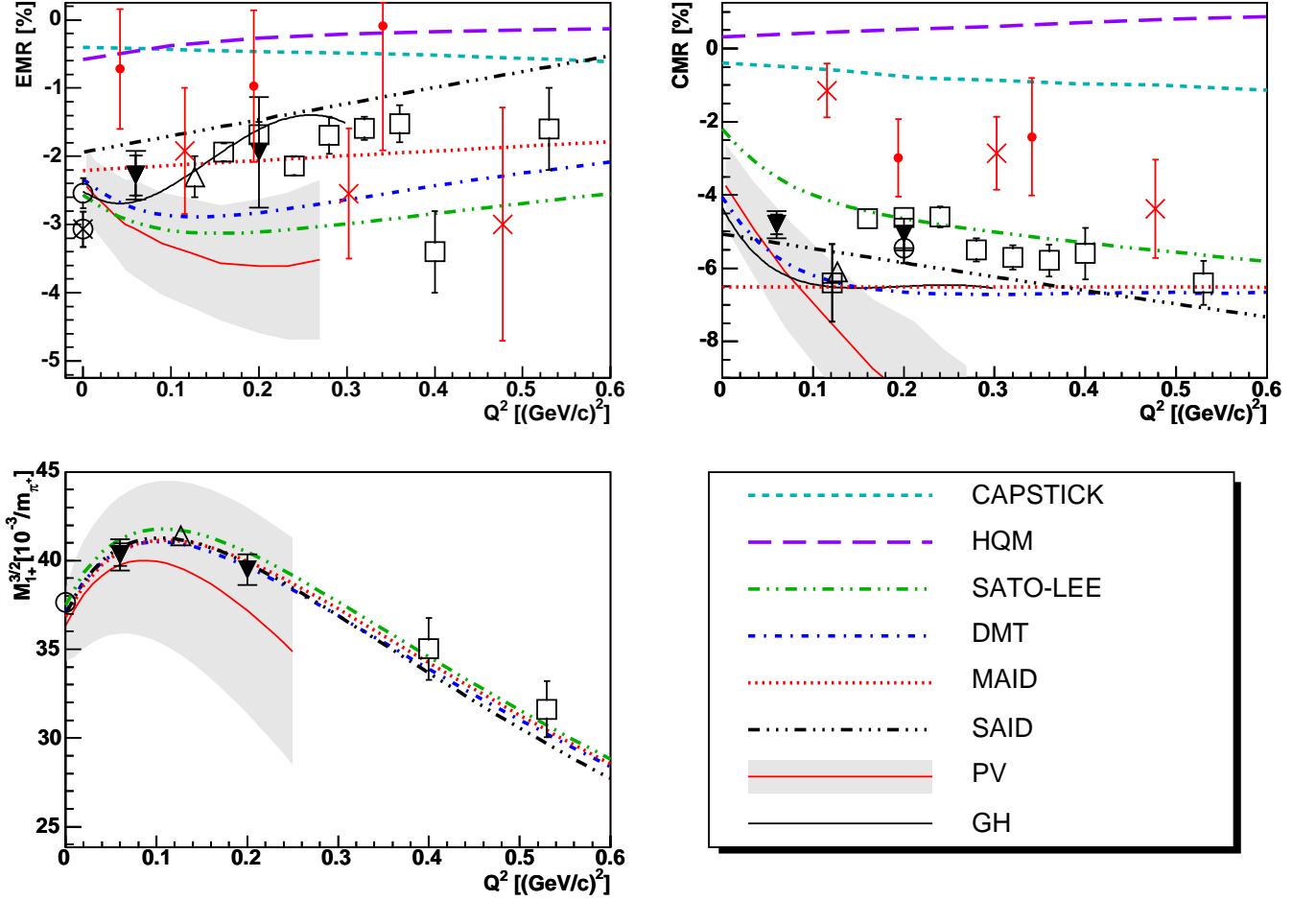}
\end{center}
\caption{\label{fig:mpole_vs_Q2}(Color online)
The low $Q^2$ dependence of the $M_{1+}$, EMR, and CMR at $W=1232$ MeV for the $\gamma^{*} p \rightarrow \Delta$ reaction. The $\blacktriangledown$ symbols are our data
points and include the experimental and model uncertainties (see
Table~\ref{table:fit:bothQ}) 
added in quadrature. The other data are the photon point
data $\bigcirc$~\cite{beck} and $\otimes$~\cite{blanpied}, CLAS
$\square$~\cite{joo,CLAS2007} (CLAS data for $0.16 \le Q^2 \le 0.36$ \gevc
are from a unitary isobar model fit and have statistical uncertainties only),  Bates $\triangle$~\cite{sparveris}, Elsner
$\bigoplus$~\cite{elsner},  
and Pospischil $\boxplus$~\cite{pospischil}.  All uncertainties are
statistical and systematic added in quadrature unless otherwise
noted.  The lattice QCD calculations with linear pion mass
extrapolations are shown as $\times$~\cite{alexandrou} and the new
calculations with small pion mass but without extrapolation are shown as
$\bullet$~\cite{alexandrou2}.  Also shown are the chiral perturbation calculations of Pascalutsa and
  Vanderhaeghen (PV) (see EFT in Fig.~\ref{fig:Q06_XS_all})~\cite{pasc} 
and Gail and Hemmert (GH) (block solid line)~\cite{gail_hemmert}.
The other curves represent the same models as in Fig.~\ref{fig:Q06_XS_all}. 
The HQM (long-dashed line)~\cite{hqm} and
Capstick (short-dashed line)~\cite{capstick_karl}  quark models have been included. }
\end{figure*}

One of the main goals of this experiment is to determine the $Q^2$
variation of the resonant multipoles.     
The new Jefferson Lab results~\cite{CLAS2007} for the EMR and CMR at $Q^2=0.200$ \gevc agree with
our results very well.  In addition, it was already shown that the
present Mainz data
agree with the previous Bates data~\cite{mertz} at $Q^2=0.127$ \gevcp.  All of
this shows that there is reasonable consistency of the results from
the different laboratories.  

Figure \ref{fig:mpole_vs_Q2} shows
the evolution of the multipoles at low $Q^2$ along with all the other
published points. Two representative
constituent quark models, the newer hypercentral quark model
(HQM)~\cite{hqm}, and an older non-relativistic calculation of Capstick~\cite{capstick_karl}, have been included (the relativistic
calculations are in even worse agreement with experiment). These
curves are representative of quark models which typically
under-predict the dominant $M_{1+}^{3/2}$ multipole by $\simeq$ 30\%
and underestimate the EMR and CMR by an order of magnitude, even
predicting the wrong sign. One solution to this problem has been to
add pionic degrees of freedom to quark models~\cite{quark_pion_1,
  quark_pion_2, quark_pion_3}.
  All of these models treat the $\Delta$ as a bound state and therefore
do not have the $\pi$N continuum (i.e. no background amplitudes) so
that cross sections are not calculated. The Sato-Lee~\cite{sato_lee}
and DMT~\cite{dmt} dynamical reaction models with pion cloud effects
bridge this gap     and  are in qualitative agreement with the $Q^{2}$
evolution of the data. These models calculate the virtual pion cloud
contribution dynamically but have an empirical parameterization of the
inner (quark) core contribution which gives them some flexibility in
these observables.  By contrast the empirical MAID~\cite{maid1} and
SAID~\cite{said} represent fits to other data with a smooth $Q^2$
dependence.

 Both the dynamical~\cite{dmt,sato_lee} 
and the phenomenological~\cite{maid1,said} models
are in qualitative
agreement with the experimental results. Nevertheless, all models exhibit
some small deficiencies
either on top or at the wings of the resonance indicating that detailed
improvements could and
should be implemented to the models description of resonant or background
amplitudes towards
accounting for these deficiencies. As a general remark one can note the
much better behavior
of the dynamical models (DMT and Sato-Lee) compared to the phenomenological ones
(MAID and SAID) as far as the
description of the $W$ evolution of the cross section is concerned (see
Fig.~\ref{fig:Q06_W}) while for Sato-Lee
the description of the fifth response is also excellent thus indicating that
the model provides
the most consistent description of the background amplitudes. One must also
point out though the
consistent description that SAID provides for the unpolarized cross
sections on top of the
resonance measurements for all $Q^2$ points.

The plotted lattice QCD results with a
linear pion mass extrapolation~\cite{alexandrou} are in general
agreement with the data for the EMR but disagree for the CMR by a wide
margin.  
This margin is bridged, though, when using a chiral extrapolation to the
physical pion mass
instead of the linear one. 
The EFT analysis of Pascalutsa and Vanderhaeghen (PV)~\cite{pasc} indicates that a linear extrapolation is close to the data
for the EMR but not for the CMR for which these extrapolated lattice
results are considerably reduced. 
The second plotted lattice QCD results were
performed with an improved method an a smaller pion mass and are
reported without any extrapolation~\cite{alexandrou2}.
It is significant that these newer results have the same sign as the data at low $Q^2$.
The general qualitative agreement of the lattice QCD calculation 
provides a direct link with the experimental evidence for deformation
to QCD.

The results of the two effective field theory
calculations~\cite{gail_hemmert,pasc} are also presented in
Fig. \ref{fig:mpole_vs_Q2}. These 
contain empirical low energy constants. For Gail and Hemmert this
includes fits to  the dominant $M_{1+}^{3/2}$ multipole for $Q^{2}\leq
0.2$ \gevc and for the EMR at the photon point ($Q^{2}$=0). In order
to achieve the good overall agreement they had to employ one  higher
order  term with another empirical constant. The EFT calculation of
Pascalutsa and Vanderhaeghen~\cite{pasc} provided a valuable estimate
of the uncertainties caused by excluding the next higher order terms
from the calculation.  While this is a very helpful start, the
uncertainties are significantly larger than the experimental uncertainties and will
have to be reduced through a proper treatment of the excluded higher
order terms.
However, these effective field theoretical (chiral) calculations that are
solidly based on QCD,
successfully account for the magnitude of the effects giving further credence
to the dominance
of the meson cloud effect.

One way to see the major role played by the pion cloud contribution to
the resonant multipoles is that for this case the expected scale for
the $Q^2$ evolution is $m_\pi^2=0.02$ GeV$^2$.  In these units the
range of the present experiment for $Q^2$ from 0.060 to 0.200 \gevc is
3 to 10 units.  Therefore it is not surprising that one should see
relatively large changes in the predicted $Q^2$ evolution of the
resonant multipoles as is shown in Fig. \ref{fig:mpole_vs_Q2}.  It is
also clear that there is significant model dependence in these predictions.


\FloatBarrier

\section{Conclusions}

The data presented here provide a precise determination of the
resonant amplitudes in the $\gamma^{*} p \rightarrow \Delta$ reaction in the
range of $Q^2=0.06$
to $Q^2=0.20$ \gevc (3 to 10 $m_{\pi}^{2}$). The experiment at the Mainz
Microtron  was carefully designed to reach the lowest possible $Q^2$  to
test effective field theory calculations and to probe the regime where
pionic effects are predicted to be a maximum and to vary
significantly~\cite{sato_lee,dmt}. The absolute cross section accuracy at
the 3\% level was
verified with several cross-checks.  The measurement of the $\sigma_{0}=
\sigma_{T}+\epsilon
\sigma_{T}, \sigma_{LT},\sigma_{TT}$ and $\sigma_{LT'}$ partial cross
sections, at center of mass energies both on and off resonance, allows for
sensitive tests of effective field theory~\cite{gail_hemmert,pasc} and
reaction model calculations~\cite{sato_lee,dmt,said,maid1}.
These partial cross sections are also important for extracting the resonant
multipoles from the data; these are used to test
 lattice calculations~\cite{alexandrou,alexandrou2} and quark models~
\cite{hqm,capstick}.  At the present time the experiments are more accurate
than both theory and model calculations.

 The chiral effective field theory predictions~\cite{gail_hemmert,pasc} agree
with our cross section data within the relatively large estimated
theoretical uncertainties due to the neglect of higher order terms. It is
clear that a quantitative comparison of these calculations and
experiment must wait until the next order calculations are performed.
The phenomenologically adjusted models Sato-Lee, DMT, SAID and MAID~
\cite{sato_lee,dmt,said,maid1}
are in good agreement with experiment when the
resonant amplitudes are adjusted to the data. This allows an accurate
extraction of the M1, E2 and C2 resonant multipoles ($M_{1+}^{3/2}, E_
{1+}^{3/2},
S_{1+}^{3/2}$) with an estimated model uncertainty  which is approximately
the same as the experimental uncertainty. This has been achieved due to the
precision of the experimental data and also because of
the dominance of the magnetic dipole amplitude $M_{1+}$; this dominance
means that differences in the background amplitudes are not significant
near resonance and that the model uncertainties in the determination of the
resonant multipoles  are comparable with with the experimental uncertainties. The
differences in the background amplitudes  have been demonstrated in our
 low $W$ data and in $\sigma_{LT'}$ for which the background 
multipoles play a more significant role. This emphasizes the need for model
builders to improve their calculations and also to present their uncertainties, as
has been done in
the EFT calculations. We have performed our own
error estimate by comparing the extracted resonance multipoles using
different models.

Comparisons of the measured resonant multipoles as a function of $Q^2$ show
reasonable agreement between experiments at different laboratories.
The non-zero values of the quadrupole
amplitudes ($E_{1+}^{3/2},S_{1+}^{3/2}$) demonstrate the existence of
non-spherical amplitudes in the nucleon
and $\Delta$  conjectured many years ago on the basis of the non-spherical
interaction between quarks~\cite{glashow}. This feature is also present in
the lattice calculations~\cite{alexandrou,alexandrou2}, thus linking the
experimental evidence for deformation directly to QCD. Unfortunately, the
uncertainties in the present calculations are large, which precludes a quantitative
comparison with experiment.
We anticipate further advances with calculations at lower quark masses
combined with improved chiral calculations which are also just in the
beginning~\cite{pasc}.
These results show qualitative agreement with the two chiral
effective field theory results~\cite{gail_hemmert,pasc}. The uncertainties
in these latter two
calculations indicate that higher order terms must be evaluated before a
quantitative comparison can be made.

Comparison with representative quark models~\cite{hqm,capstick}
shows that they are not close to the data
indicating a deficiency of the underlying physics description while
demonstrating that the color hyperfine
interaction is inadequate to explain the effect,  at least at large
distances. Our present
understanding is that the long range (low $Q^2$) region is dominated
by  the spontaneous breaking of chiral symmetry in QCD which results
in  non-spherical pion emission and absorption from the nucleon and $
\Delta$~\cite{athens2006,mit2004,nstar2001,cnp,amb}.

Even though experiments are ahead of theory at the present time,
future experiments can add to the current understanding by measuring in the
$n\pi^+$ and $\gamma$ channels and by utilizing polarized targets and
polarimeters.  New data for the $\gamma$ channel from Mainz are under
analysis
and should be published soon~\cite{sparveris_vcs}.  Additional low $W$ data
would also give a better handle on the background amplitudes.

\FloatBarrier

\begin{acknowledgments}
We thank
 L. Tiator,  D. Drechsel, T.-S. H. Lee, V. Pascalutsa, M. Vanderhaeghen,
T. Gail, T. Hemmert and L. C. Smith for their assistance with valuable discussions and for sharing their unpublished work.  
This work is supported at Mainz by the Sonderforschungsbereich 443 of
 the Deutsche Forschungsgemeinschaft (DFG),  
U. Athens by the Program PYTHAGORAS of the Greek ministry of 
Education (co-funded by the European Social Fund and National Resources),
and at MIT by the  U.S. DOE under Grant No. DE-FG02-94ER40818.

\end{acknowledgments}

\appendix*
\section{Data Tables \label{app:data}}

\begin{table}[htb]
\begin{ruledtabular}
\begin{tabular}{cccc}
$W$ & $\theta_{pq}^*$ & $\phi_{pq}^{*}$ & $\sigma$ \\
$[\rm MeV]$  & $[^\circ]$ & $[^\circ]$ &
[$\mu$b/sr]\\
\hline
1221&	0&	---&	12.35 $\pm$ 0.09 $\pm$ 0.38\\
1221&	24&	0&      11.65 $\pm$ 0.06 $\pm$ 0.36\\
1221&	24&	90&     18.67 $\pm$ 0.09 $\pm$ 0.58\\
1221&	24&	180&    15.39 $\pm$ 0.07 $\pm$ 0.48\\
1221&	37&	32&     15.67 $\pm$ 0.12 $\pm$ 0.52\\
1221&	37&	134&    23.38 $\pm$ 0.12 $\pm$ 0.73\\
1221&	37&	180&    17.87 $\pm$ 0.08 $\pm$ 0.56\\
\hline
1155 & 26 & 0 & 5.57 $\pm$ 0.05 $\pm$ 0.20\\
1155 & 26 & 180 & 6.38 $\pm$ 0.04 $\pm$ 0.23\\
\hline
1125&	0&	---& 2.40 $\pm$ 0.02 $\pm$ 0.09\\
1155&	0&	---& 5.48 $\pm$ 0.06 $\pm$ 0.20\\
1185&	0&	---& 10.27 $\pm$ 0.10 $\pm$ 0.39\\
1205&	0&	---& 12.58 $\pm$ 0.11 $\pm$ 0.47\\
1225&	0&	---& 10.88 $\pm$ 0.10 $\pm$ 0.41\\
1245&	0&	---& 7.21 $\pm$ 0.09 $\pm$ 0.27\\
1275&	0&	---& 3.17 $\pm$ 0.04 $\pm$ 0.12\\
1300&	0&	---& 1.48 $\pm$ 0.02 $\pm$ 0.06\\
\end{tabular}
\end{ruledtabular}
\caption{\label{table:results:lowQ:XS}$Q^2=0.060$ \gevc cross
  sections.  The first uncertainty is the statistical uncertainty and the second
  is the systematic uncertainty. The helicity
dependent cross sections are shown in Table
\ref{table:results:sigs_Q06}.  
The uncertainties are statistically
dominated for those results.  The third set of results are from the
background amplitude test.  The lower set of results is the $W$ parallel cross
section scan.  }
\end{table}

\begin{table}[htb]
\begin{ruledtabular}
\begin{tabular}{ccccc}
$W$ & $Q^2$ & $\theta_{pq}^*$ & $\sigma$ & $\sigma$\\
 ${\rm [MeV]}$ & [\gevc] & $[^\circ]$ &       & [$\mu$b/sr]\\
\hline
1221 & 0.060 & 24.0 & $\sigma_0$ & $16.10 \pm 0.17 \pm 0.44$\\
1221 & 0.060 & 24.0 & $\sigma_{TT}$ & $-3.30 \pm 0.22 \pm 0.10$\\
1221 & 0.060 & 24.0 & $\sigma_{LT}$ & $1.12 \pm 0.09 \pm 0.04$ \\
1221 & 0.060 & 37.0 & $\sigma_0$ & $21.02 \pm 0.31 \pm 0.58$ \\
1221 & 0.060 & 37.0 & $\sigma_{TT}$ & $-7.99 \pm 0.63 \pm 0.20$\\
1221 & 0.060 & 37.0 & $\sigma_{LT}$ & $1.85 \pm 0.13 \pm 0.05$\\
\hline
1155 & 0.060 & 26.0 & $\sigma_{LT}$ &  $0.22 \pm 0.06 \pm 0.07$\\
1155 & 0.060 & 26.0 & $\sigma_{0}+\epsilon\sigma_{TT}$ &  $5.97 \pm 0.11 \pm 0.10$ \\
\hline
1221 & 0.060 & 24.0 & $\sigma_{LT'}$ & $1.23 \pm 0.18 \pm 0.04$ \\
1221 & 0.060 & 37.0 & $\sigma_{LT'}$ & $1.59 \pm 0.35 \pm 0.06$ \\
\end{tabular}
\end{ruledtabular}
\caption{\label{table:results:sigs_Q06} Summary of the extracted
  values for $\sigma_0$, $\sigma_{TT}$, $\sigma_{LT}$, and
  $\sigma_{LT'}$.  The uncertainties in the cross section are the 
  statistical and systematic uncertainty respectively.  See text for
  details of the uncertainty estimation procedure.
}
\end{table}

\begin{table}[htb]
\begin{ruledtabular}
\begin{tabular}{cccc}
$W$ & $\theta_{pq}^*$ & $\phi_{pq}^{*}$ &  $\sigma$ \\
$[\rm MeV]$  & $[^\circ]$ & $[^\circ]$ & 
[$\mu$b/sr] \\
\hline
1140&	58.6&	45&	 $6.93 \pm 0.08 \pm 0.26$\\
1140&	58.6&	135&	 $5.53 \pm 0.04 \pm 0.21$\\
\hline
1221&   30 & 90 &   $22.61 \pm 0.16 \pm 0.80$\\
1221&   43 & 135 &  $26.97 \pm 0.21 \pm 0.95$\\
1221&   63 & 150 &  $29.51 \pm 0.23 \pm 1.04$\\
\hline
1205& 0 & --- & $13.92 \pm 0.11 \pm 0.54$\\
1232& 0 & --- & $10.89 \pm 0.09 \pm 0.36$\\
\end{tabular}
\end{ruledtabular}
\caption{\label{table:results:Q127:XS}$Q^2=0.127$ \gevc results.  The
  first uncertainty is statistical and the second is the systematic.
 The helicity dependent results are in Table \ref{table:results:sigs_Q127}.
}
\end{table}

\begin{table}[htb]
\begin{ruledtabular}
\begin{tabular}{ccccc}
$W$ & $Q^2$ & $\theta_{pq}^*$ & $\sigma$ & $\sigma$\\
 ${\rm [MeV]}$ & [\gevc] & $[^\circ]$ &       & [$\mu$b/sr]\\
\hline
1140  & 0.127 & 58.6 & $\sigma_0$ & $6.23 \pm 0.12 \pm 0.12$\\
1140  & 0.127 & 58.6 & $\sigma_{LT}$ & $-0.58 \pm 0.10 \pm 0.10$\\
\hline
1140  & 0.127 & 58.6 & $\sigma_{LT'}$ & $0.94 \pm 0.16 \pm 0.04$\\
1221  & 0.127 & 30.0 & $\sigma_{LT'}$ & $1.84 \pm 0.28 \pm 0.07$\\
1221  & 0.127 & 43.0 & $\sigma_{LT'}$ & $2.80 \pm 0.51 \pm 0.10$\\
1221  & 0.127 & 63.0 & $\sigma_{LT'}$ & $1.25 \pm 0.78 \pm 0.05$\\
\end{tabular}
\end{ruledtabular}
\caption{\label{table:results:sigs_Q127} Summary of the extracted
  values for $\sigma_0$, $\sigma_{LT}$, and
  $\sigma_{LT'}$ at $Q^2=0.127$ \gevcp.   The uncertainties in the cross section are the 
  statistical and systematic uncertainty respectively.
}
\end{table}

\begin{table}[htb]
\begin{ruledtabular}
\begin{tabular}{cccc}
$W$ & $\theta_{pq}^*$ & $\phi_{pq}^{*}$ &  $\sigma$ \\
$[MeV]$  & $[^\circ]$ & $[^\circ]$ &
[$\mu$b/sr] \\
\hline
1221 &    0 &    --- &    $12.29 \pm 0.10 \pm 0.51$  \\
1221 &    27 &    0 &    $12.25 \pm 0.13 \pm 0.46$  \\
1221 &    27 &    90 &    $18.28 \pm 0.16 \pm 0.65$  \\
1221 &    27 &    180 &    $16.94 \pm 0.15 \pm 0.58$  \\
1221 &    33 &    0 &    $12.71 \pm 0.11 \pm 0.43$  \\
1221 &    33 &    90 &    $21.73 \pm 0.18 \pm 0.76$  \\
1221 &    33 &    180 &    $18.09 \pm 0.14 \pm 0.63$  \\
1221 &    40 &    0 &    $13.72 \pm 0.14 \pm 0.52$  \\
1221 &    40 &    90 &    $26.46 \pm 0.21 \pm 0.92$  \\
1221 &    40 &    180 &    $19.28 \pm 0.15 \pm 0.67$  \\
1221 &    57 &    38 &    $23.20 \pm 0.20 \pm 0.81$  \\
1221 &    57 &    142 &    $27.86 \pm 0.21 \pm 0.97$  \\
1221 &    57 &    180 &    $22.75 \pm 0.19 \pm 0.77$  \\
\hline
1125 &	0 &	--- &	$2.32 \pm 0.02 \pm 0.08$\\
1155 &	0 &	--- &	$5.72 \pm 0.04 \pm 0.19$\\
1185 &	0 &	--- &	$10.66 \pm 0.08 \pm 0.36$\\
1205 &	0 &	--- &	$12.96 \pm 0.09 \pm 0.44$\\
1225 &	0 &	--- &	$11.62 \pm 0.08 \pm 0.39$\\
1245 &	0 &	--- &	$7.84 \pm 0.06 \pm 0.26$\\
1275 &	0 &	--- &	$3.71 \pm 0.04 \pm 0.12$\\
\end{tabular}
\end{ruledtabular}
\caption{\label{table:results:Q20:XS}$Q^2=0.200$ \gevc 
cross section results. The uncertainties correspond to the statistical and
the systematic uncertainties respectively.  The lower set of results is the $W$ parallel cross
section scan.}
\end{table}

\begin{table}[htb]
\begin{ruledtabular}
\begin{tabular}{ccccc}
$W$ & $Q^2$ & $\theta_{pq}^*$ & $\sigma$ & $\sigma$  \\
 ${\rm [MeV]}$ & [\gevc] & $[^\circ]$ &       & [$\mu$b/sr] \\
\hline
1221 & 0.20 & 27.0 & $\sigma_0$ & $16.44 \pm 0.19 \pm 0.65$  \\
1221 & 0.20 & 27.0 & $\sigma_{TT}$ & $-2.99 \pm 0.15 \pm 0.32$  \\
1221 & 0.20 & 27.0 & $\sigma_{LT}$ & $1.66 \pm 0.07 \pm 0.13$  \\
1221 & 0.20 & 33.0 & $\sigma_0$ & $18.56 \pm 0.21 \pm 0.68$   \\
1221 & 0.20 & 33.0 & $\sigma_{TT}$ & $-5.13 \pm 0.16  \pm 0.31$  \\
1221 & 0.20 & 33.0 & $\sigma_{LT}$ & $1.90 \pm 0.06  \pm 0.12$  \\
1221 & 0.20 & 40.0 & $\sigma_0$ & $21.48 \pm 0.24  \pm 0.99$  \\
1221 & 0.20 & 40.0 & $\sigma_{TT}$ & $-8.08 \pm 0.19  \pm 0.59$  \\
1221 & 0.20 & 40.0 & $\sigma_{LT}$ & $1.97 \pm 0.07  \pm 0.15$  \\
1221 & 0.20 & 57.0 & $\sigma_0$ & $27.36 \pm 0.49  \pm 1.14$  \\
1221 & 0.20 & 57.0 & $\sigma_{TT}$ & $-12.28 \pm 0.66  \pm 1.06$  \\
1221 & 0.20 & 57.0 & $\sigma_{LT}$ & $2.10 \pm 0.12  \pm 0.24$  \\
\hline
1221 & 0.20 & 33.0 & $\sigma_{LT'}$ & $2.10 \pm 0.22  \pm 0.35$  \\
1221 & 0.20 & 57.0 & $\sigma_{LT'}$ & $2.24 \pm 0.26  \pm 0.41$  \\
\end{tabular}
\end{ruledtabular}
\caption{\label{table:results:sigs_Q20} Extracted
  values for $\sigma_0$, $\sigma_{TT}$, $\sigma_{LT}$, and
  $\sigma_{LT'}$ at $Q^2=0.200$ \gevcp. The uncertainties correspond to the statistical and
the systematic uncertainties respectively.}
\end{table}

\begin{table}[htb]
\begin{ruledtabular}
\begin{tabular}{cccc}
$W$ & $\theta_{pq}^*$ & $\phi_{pq}^{*}$ &  $\sigma$\\
$[\rm MeV]$  & $[^\circ]$ & $[^\circ]$ &
[$\mu$b/sr]\\
\hline
1205 &  0 &	--- &	$11.25 \pm 0.18 \pm 0.54$\\
\end{tabular}
\end{ruledtabular}
\caption{\label{table:results:Q30:XS}$Q^2=0.300$ \gevc results.  The
  first uncertainty is statistical and the second is systematic. }
\end{table}

\FloatBarrier

\end{document}